\documentclass[%
 reprint,
 amsmath,amssymb,superscriptaddress,
prl
]{revtex4-1}
\usepackage{xparse}

\usepackage{mathtools}
\usepackage{appendix}
\usepackage{graphicx}% Include figure files
\usepackage{dcolumn}% Align table columns on decimal point
\usepackage{bm}% bold math
\usepackage{hyperref}% add hypertext capabilities
\usepackage{xcolor}
\usepackage[normalem]{ulem}
\newcommand{\eps}{\ensuremath{\varepsilon}}

\newcommand{\diff}[1]{\ensuremath{\text{d} #1}}

\newcommand{\Tr}{\text{Tr}}

\newcommand{\R}{\mathtt{R}}

\newcommand{\mdet}{\mathtt{M}}
\newcommand{\wdet}{\mathtt{W}}

\usepackage{bbold}

\begin{document}

\title{Many-body level statistics of single-particle quantum chaos}

\author{Yunxiang Liao}
\author{Amit Vikram}
\author{Victor Galitski}
\affiliation{Joint Quantum Institute, University of Maryland, College Park, MD 20742, USA.}
\affiliation{Condensed Matter Theory Center, Department of Physics, University of Maryland, College Park, MD 20742, USA} 
%\date{\today}
\begin{abstract}
		We consider a non-interacting many-fermion system populating levels of a unitary random matrix ensemble (equivalent to the $q=2$ complex Sachdev-Ye-Kitaev model) -- a generic model of single-particle quantum chaos. We study the corresponding {\em many-particle level statistics} by calculating the spectral form factor  analytically using algebraic methods of random matrix theory, and match it with an exact numerical simulation. Despite the integrability of the theory, the many-body spectral rigidity  is found to have a surprisingly rich landscape. In particular,  we find a residual repulsion of  distant many-body levels  stemming from single-particle chaos, together with islands of level attraction. These results are encoded in an exponential ramp in the spectral form-factor, which we show to be a universal feature of non-ergodic many-fermion systems embedded in a chaotic medium.  
\end{abstract}	

\maketitle

There has been growing recent  interest in the foundational questions of statistical mechanics, from the eigenstate thermalization hypothesis~\cite{ETH-review,srednicki1994chaos,srednicki1999,rigol2008,srednicki2019bounds} 
to many-body localization~\cite{BAA1,BAA2,BAA3,Imbrie,Huse2007,Review-Huse,Review-Serbyn}.
%\cite{Huse2010,Imbrie,MBL1D-0,MBL1D-1,MBL1D-2,MBL1D-3,MBL1D-4,MBL1D-5,MBLexp1D,MBLexp2D,DeMarco,Monroe}.
These questions are intimately related to the notion of quantum chaos~\cite{ReviewChaos}. While intuitive, quantum chaos is not easy to define. The usual approach is to associate quantum-chaotic, ergodic systems with energy spectra exhibiting  Wigner-Dyson (WD) level statistics \cite{BG,BGS,refRMTMehta,dyson,wigner2012}. On the other hand, integrable (including localized) non-ergodic systems are expected to exhibit Poisson level statistics - lack of any correlation between the levels \cite{refBerryPoisson}. A lot is known about single-particle quantum chaos, which has been explored for a variety of systems from chaotic billiards~\cite{BGS,mcdonald,casati,berryB,billiardootc} to disordered metals~\cite{Eliashberg,Efetov,Efetov2,Altshuler,Zirnbauer,Kravtsov,Kamenev,Kamenev2}, where the WD level statistics has indeed been seen using numerical and analytical theoretical techniques, as well as in experiment~\cite{WDexpt, WDexpt2, WDexpt3, WDexpt4, WDexpt5}. 

Many-body quantum chaos~\cite{abergchaos,Prosen-chaos,Prosen-chaos2,Muller-chaos,Jacquod,Green,gornyimanybodychaos,Cotler-2017} is a more difficult concept to both define and study. The structure of many-body energy levels is very fine, with nearest neighbor level spacing
inversely proportional to the Hilbert space size, which is exponential in system size, $N$, which itself is usually an astronomically large number in most many-body systems
of interest. Furthermore, there are confusingly two types of ``quantum chaos'' that may be present in a many-body system. Consider for example a weakly disordered system in three dimensions. A single  quantum particle moving in this random potential will  exhibit WD level statistics -- a classic result of Altshuler and Shklovksii~\cite{Altshuler}, which is the hallmark of single-particle quantum chaos.  If we embed a  non-interacting  $N$-particle (e.g., $N$-electron) system in this random medium, it will never thermalize due to lack of interactions. This Fermi gas is an integrable system, which is not associated with many-body quantum chaos.  On the other hand, a generic interacting many-body Fermi system is expected to thermalize - i.e., exhibit an ergodic, many-body quantum-chaotic behavior (which presumably should exist with or without an initial disordered potential). Of great interest is the open problem of a transition or perhaps energy-dependent crossover from single-particle to many-body quantum chaos  in the distribution of energy levels of such a system.  

%Among the existing results are \textcolor{blue}{Ref.~\cite{refBerryPoisson}} by Berry and \textcolor{blue}{[Ref]} by  Zurek, who have considered the non-interacting limit and provided qualitative arguments that the many-body level statistics is Poisson-like. Also,
Several works~\cite{SYK-Stanford,SYK-Cotler, SYK-Altland,SYK-Altland-2,SYK-You,SYK-garcia,SYK-Garcia-2,deformed-SYK,SYK-Liu,SYK2-2019,SYK2-2020, SYK2-Bohigas, Sachdev} considered the many-body level statistics of a family of the Sachdev-Ye-Kitaev (SYK) and related models using a combination of field theory technique and numerical simulations. However, explicit analytical results for level correlators are lacking even in the non-interacting case (or equivalently, SYK-2). In this work, we calculate the many-body level statistics of a single-particle quantum chaotic model. We show that the corresponding spectral form factor, which encodes the 2-level statistics of the many-body system, is not pure Poisson and retains rich structure descending from single-particle chaos of the underlying model.

\textit{Model}--- As a simple model exhibiting quantum chaos, we choose a Gaussian Unitary ensemble (GUE)~\cite{refRMTMehta,Kamenev-GUE} of $N\times N$ Hermitian single-particle Hamiltonians, $\hat{h}$, following the distribution function:
\begin{equation}
\label{eq:P_h}
 \begin{aligned}
 	P(\hat{h}) =2^{N(N-1)/2} \left( \frac{N}{2\pi}\right)^{N^2/2}  \exp \left[ -\frac{N}{2}\Tr \left( \hat{h}^2 \right)  \right],
 \end{aligned}
\end{equation}
with the local level statistics of $\hat{h}$ falling into the unitary WD class~\cite{refHaake, refRMTMehta}. Populating these single-particle energy levels with fermions ($\hat{f}_i,\hat{f}_i^\dagger$) with a chemical potential $\mu$ then defines the many-body Hamiltonian,
\begin{equation}
\hat{H} = \sum_{i,j}\hat{f}_i^\dagger (h_{ij}-\mu\delta_{ij}) \hat{f}_j.
\label{eq:cSYK}
\end{equation}
%\textcolor{blue}{We can express this in terms of the eigenvalues $\eps_i$ of $h_{ij}$ with the help of a unitary transformation, giving $\hat{H} = \sum_{i}\hat{{f'}}^\dagger_i\hat{{f'}}_i(\eps_i-\mu)$, which reveals the integrable nature of the system. Specifically, the particle numbers $\hat{n}_i = \hat{{f'}}^\dagger_i\hat{{f'}}_i$ are constants of the motion.}
This is an integrable model, and the particle number at each single-particle level is a constant of the motion.

In general, for a statistical ensemble of Hamiltonians, we can define a representative 2-point spectral form factor (SFF)~\cite{BerrySFF, refHaake},
\begin{equation}
    K(t) \equiv \langle\lvert Z(it)\rvert^2\rangle = \left\langle\sum_{n,m}e^{i(E_m-E_n)t}\right\rangle, \label{eq:SFF_FT}
\end{equation}
where $Z(it) \equiv \Tr(e^{-i\hat{H}t})= \sum_{n}e^{-iE_n t}$, with $E_n$ being the eigenvalues of $\hat{H}$, and
the angular bracket represents ensemble averaging.
It immediately follows that $0\leq \lvert K(t)\rvert \leq L^2$ with $K(0) = L^2$, and $K(\infty) = L$ if degeneracies are statistically insignificant in the ensemble, where $L$ is the Hilbert space size i.e. number of energy levels of the system.

The SFF is essentially a Fourier transform of the joint two-level distribution function (see also Eq.~\eqref{eq:rho2_def}). For an ensemble with Poisson statistics (independently distributed energy levels), $K(t)$ decays from $L^2$ at $t=0$, gradually approaching $L$ at a time scale much smaller than the inverse mean level separation. 
However, Hamiltonians obeying WD statistics are characterized by level repulsion at a scale $\Delta$ corresponding to the typical level spacing. This results in an SFF that ``slopes'' down below $L$ up to around a dip time $t \sim t_d$, where it reaches a minimum, then grows in an approximately linear ``ramp'', abruptly reaching a $K(t)=L$ ``plateau'' for $t\geq t_\ast$. We will call $t_\ast$ the plateau time. The ramp and plateau have their origins in the Fourier transform of the level repulsion component of the distribution, which implies that $t_\ast\sim 1/\Delta$. To the extent that the level repulsion is given by the WD universality classes, the ramp and plateau are also universal features of quantum chaotic systems (see e.g. Refs.\cite{refRMTMehta, refHaake, Cotler-2017,refRampPlateau2}).

\textit{Results}--- For the system given by Eq.~\eqref{eq:cSYK}, which is our primary concern in this paper, we have $N$ single-particle levels, and $L=2^N$. We find three approximate expressions that closely describe the SFF in different regions, in the large $N$ limit, i.e.
\begin{equation}
K(t) \approx \begin{cases}
K_1(t), &\ 0 < t \ll O(1), \\
K_2(t),  &\ O(1) \ll t \ll  O\left(N/\log_2 N\right),\\
K_3(t), &\sqrt{2}N < t < \infty,
\end{cases}
\label{eq:SFF_result}
\end{equation}
where
\begin{subequations}
\begin{align}
& \begin{aligned}
K_1(t) = L^2 \cos^{2N}\left(\frac{\mu t}{2}\right) \exp&\left[ N\left(\frac{J_1(2t)}{t}-1\right)\cos(\mu t) \right],
\end{aligned} \label{eq:K1}\\
& \begin{aligned} K_2(t) =  \left( \frac{N}{8} e^{\gamma_E}\right)^{t/4} \exp\left[ N \dfrac{J_1(2t)}{t}\cos(\mu t)\right],\end{aligned} \label{eq:K2}\\
& \begin{aligned}
K_3(t) = L\exp\left[-\frac{(4N^2-t^2)^\frac{3}{2}}{12\pi Nt}\Theta(2N-t)\right],
\end{aligned} \label{eq:K3}
\end{align}
\end{subequations}
with $J_1(z)$ being the Bessel function of the first kind, $\Theta(x)$ the unit step function, and $\gamma_E = 0.577...$ the Euler-Mascheroni constant.
$K_1(t)$ describes the initial downward slope region; $K_2(t)$ is related to the transition from an oscillatory region up to $t\sim O((N/\ln N)^\frac{2}{5})$ to an exponential beginning of the ramp; and $K_3(t)$ gives the late-time ramp approaching the plateau.
\begin{figure}[t!]
\centering
\includegraphics[scale=0.3]{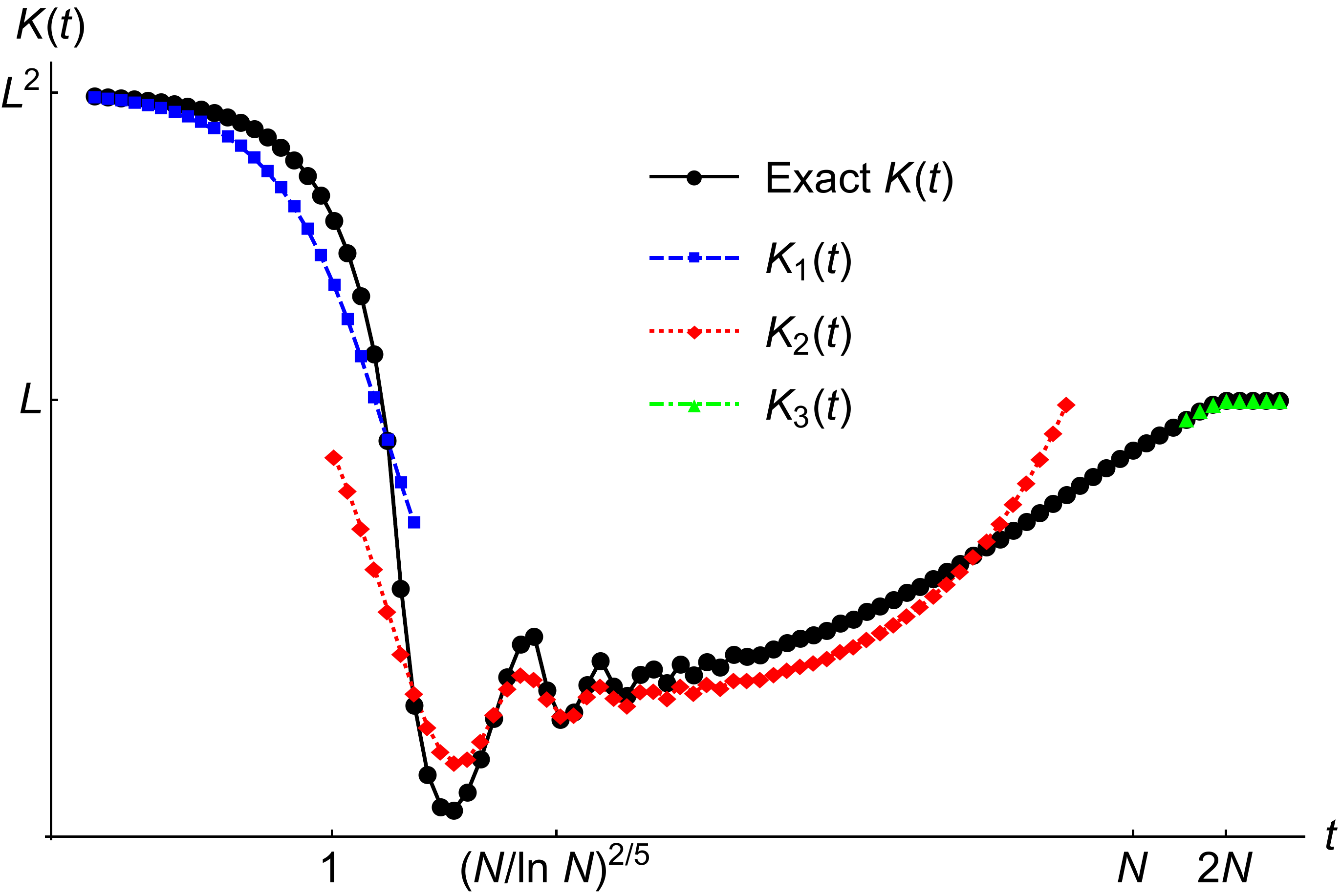}
\caption{
A log-log plot of the many-body SFF of the non-interacting GUE model (Eqs.~\eqref{eq:P_h} and~\eqref{eq:cSYK}) computed for $N = 400$ levels with chemical potential $\mu = 0$: numerically from Eqs.~\eqref{eq:SFF_det},~\eqref{eq:wdef} (``exact''), compared with the three approximate expressions $K_{1,2,3}(t)$ given by Eqs.~\eqref{eq:K1},~\eqref{eq:K2} and~\eqref{eq:K3}, respectively.
} 
\label{fig:SFF_MB}
\end{figure}
These approximations are illustrated in Fig.~\eqref{fig:SFF_MB}, where they are compared with a numerical calculation based on Eqs.~\eqref{eq:SFF_det}
%~\eqref{eq:mdef}
and~\eqref{eq:wdef} discussed later in this paper. We will now sketch some important steps in the derivation of these expressions, relegating the full details to the supplementary material~\cite{Sup}.

\textit{Details of calculation} ---
From the definition of the SFF (Eq.~\eqref{eq:SFF_FT}), it is straightforward to show that for an ensemble of Hamiltonians described by Eqs.~\eqref{eq:P_h} and~\eqref{eq:cSYK}, $K(t)$ is given by
%\diff\eps_1 &...\diff\eps_N
\begin{align}
\begin{aligned}
    K(t) 
    =
    2^N \int\diff\eps_1...\diff\eps_N P(\eps_1,...,\eps_N)\prod_{k=1}^N\left[1+\cos\left((\eps_k-\mu) t\right)\right]
%    =
%    \left\langle
%    \sum_{\alpha,\beta=1}^{L} 
%    \prod_{k=1}^{N}
%    e^{i(n_k^\alpha-n_k^\beta)\xi_k t}
%    \right\rangle
 %  \\
 %   =&
 %   2^N \int\diff\eps_1...\diff\eps_N P(\eps_1,...,\eps_N)\prod_{k=1}^N\left(1+\cos(\xi_k t)\right). \label{eq:FTcos}
\end{aligned}    
\end{align}
Using the correlation function of the single-particle levels derived in Ref.~\cite{refRMTMehta}, we find that, in the large $N$ limit, the SFF can be expressed as
\begin{align}\label{eq:KT-1}
\begin{aligned}
	K (t)
	=\,&
	2^N 
	\exp 
	\left\lbrace 
	N \dfrac{J_1(2t)}{t}\cos(\mu t)
	+
	A_0(t)
	\right.
	\\
	&\left.
	+
	2
	\sum_{p=1}^{N}
	A_p(t)  (-1)^p
	\dfrac{\sin \left(  \pi t p/2  \right)   }{ \pi t p/2  }
	\cos(p\mu t)\right\rbrace .
	\end{aligned}
	\end{align}
	Here 
	$A_p(t)$ is given by
	\begin{align}\label{eq:Ap}
	\begin{aligned}
	A_p(t)
	\equiv &
	-
	N
	\sum_{n=2}^{N}  \frac{1}{n} \frac{1}{2^n} 
	\sum_{\sum_{i=1}^{n} \zeta_i =p} 
	\left[ 1-\dfrac{t}{2N}   s(\left\lbrace \zeta_i\right\rbrace )  \right] 
	\\
	&\times
	\Theta 	\left[1 - \dfrac{t}{2N}  s(\left\lbrace \zeta_i\right\rbrace) \right],
	\end{aligned}
	\end{align}	
for any integer $p \geq 0$.	
$\sum_{\sum_{i=1}^{n} \zeta_i =p} $ represents the summation over all configurations of $\left\lbrace \zeta_i=\pm 1 \right\rbrace_{i=1}^{n}$, 
%with $i$ running over $1$ to $n$, 
obeying the constraint $\sum_{i=1}^{n} \zeta_i =p$. 
$s(\left\lbrace \zeta_i \right\rbrace )$ is the difference between the maximum and the minimum elements of the sequence $\left\lbrace 0,\sum_{i=1}^{j}\zeta_i\right\rbrace_{j=1}^{n-1}$. 
%	\textcolor{blue}{and $A_p(t)$ is defined in the supplementary material [Eq.~(S30)].}
	
We first focus on the regime of $t \ll O(N)$ in which $A_p(t)$ is given approximately by the sum of a constant and a linear-in-$t$ term. For sufficiently small $t\ll O(1)$, we can further approximate $A_p(t)$ by the constant and the factor $\frac{\sin \left(\pi t p/2\right)}{\pi t p/2}$ by $1$, obtaining a rapid decay that corresponds to the `slope' region of the SFF described by $K_1(t)$ (Eq.~\eqref{eq:K1}).
By contrast, for  $t \gg O(1)$,  $A_{p\geq 1}(t)$ is of the order of or smaller compared with $A_0(t)$.
%\approx \frac{t}{4}\left(\ln\frac{N}{8} + \gamma_E\right)-N \ln 2$. \sout{which assumes the form}
%\begin{equation}
%    A_0(t) \approx \frac{t}{4}\left(\ln\frac{N}{8} + \gamma_E\right)-N %\ln 2.
%\end{equation}
As a result, the last term in the exponent of $K(t)$ in Eq.~\eqref{eq:KT-1}, a summation of the highly oscillating functions, can be ignored, leading to $K_2(t)$ (Eq.~\eqref{eq:K2}). 
%\textcolor{blue}{A $t \ln N$ contribution in the exponent of the SFF has also been obtained by considering the zero modes contribution in the path integral formulation for $q=2$ SYK model~\cite{Brian}}.
For $t\ll O((N/\log N)^\frac{2}{5})$, the oscillatory Bessel function term in the exponent dominates, and the SFF continues to decay.
On the other hand, for $t\gg O((N/\ln N)^\frac{2}{5})$, $A_0(t)$ dominates, and the SFF shows a ramp given by $(Ne^{\gamma_E}/8)^{t/4}$.
While it is not easy to determine the upper limit of validity of $K_2(t)$ (Eq.~\eqref{eq:K2}), we can estimate an upper bound for the range of validity by noting that we must constrain $t \ll N/\log_2 N$ in this expression to avoid violating the condition $K(t) \leq L^2$.

For $t = O(N)$, it is difficult to derive the explicit expression of $K(t)$ from Eq.~\eqref{eq:KT-1}. We instead start from the following expression, obtained using the well known technique of expressing $P(\eps_1,...,\eps_N)$ as a determinant of the Hermite polynomials $H_n(x)$ ~\cite{refRMTMehta},
\begin{equation}
K(t) = 2^N \det\left[\delta_{jk}+\mdet_{jk}\left(t\right)\right]_{j,k=1,...,N},\label{eq:SFF_det}
\end{equation}
where
\begin{equation}
\begin{aligned}
&\mdet_{jk}(t) = \wdet_{jk}\left(\frac{t^2}{N}\right) \begin{cases}
(-1)^{\frac{j-k}{2}}\cos(\mu t), & j-k \text{ is even}, \\
(-1)^{\frac{j-k-1}{2}}\sin(\mu t), & j-k \text{ is odd},
\end{cases}
%\nonumber
%\end{equation}
%and
%\begin{equation}
\\
&\wdet_{j\geq k}(\tau) = \sqrt{\frac{(k-1)!}{(j-1)!}}\tau^\frac{j-k}{2} e^{-\frac{\tau}{2}} L_{k-1}^{j-k}(\tau),
\label{eq:wdef}
\end{aligned}
\end{equation}
with $\wdet_{jk} = \wdet_{kj}$, and $L_n^\alpha(x)$ are the Laguerre polynomials.

To simplify this expression further, we require an approximate asymptotic expression for $L_n^\alpha(x)$ for large $n$, with large or small $\alpha$. For this purpose, we use a modification~\cite{Sup} of a standard result given in Ref.~\cite{refErdelyiLaguerre}. One principal consequence of using this expression is that each $\mdet_{jk}(t)$ acquires a cutoff in $t$ above which it vanishes:
\begin{equation}
\mdet_{jk}(t) \propto \Theta\left(2N(j+k)-t^2\right).
\label{eq:detm_approx}
\end{equation}
Note that the maximum value of $j+k$ is $2N$, so the $\Theta$-function ensures that all the $\mdet_{jk}(t)$ vanish for $t>2N$, showing that the plateau occurs at $t_\ast =2N$, which is also the plateau time of the corresponding single-particle SFF~\cite{Cotler-2017, refRampPlateau2}.

We can further use the matrix relation $\det A = e^{\Tr\ln A}$ in Eq.~\eqref{eq:SFF_det}, and expand the exponent in powers $\mdet^n$ of $\mdet$. Due to the oscillatory nature of the polynomials, only terms with even $n$ contribute. The $n=2$ term is simple to evaluate, and gives a monotonically increasing upper bound for $K(t)$ for $t=O(N)$~\cite{Sup}, whose form is given by $K_3(t)$ in Eq.~\eqref{eq:K3} for $t>\sqrt{2}N$. It is also seen numerically that the $n>2$ terms appear to be negligible in this region, and $K_3(t)$ approximates $K(t)$ fairly closely.

\textit{Discussion}---%TENTATIVE, NEEDS CHECKING
To study the level statistics  with the help of $K(t)$, we can take the Fourier transform of Eq.~\eqref{eq:SFF_FT} to get the joint density function of two many-body levels with separation $S$, summed over the entire spectrum,
\begin{equation}
    \tilde{\R}_2(S) = \int_{-\infty}^{\infty} \frac{\diff t}{2\pi} K(\lvert t\rvert)e^{-iSt}-L\ \delta(S),
\label{eq:rho2_def}
\end{equation}
where in the second term, we have subtracted off the contribution from when the two levels are identical.
We will now set $\mu = 0$ as it simplifies our arguments without altering their essential content (as the ramp and plateau are $\mu$-independent). For $S \gg 1$, only the small-$t$ behavior of $K(t)$ is relevant. We therefore use the expression $K_1(t)$ from Eq.~\eqref{eq:K1} with $\mu = 0$. Expanding to leading order in $t$, we have $K_1(t) \approx L^2 \exp\left(-\frac{N}{2}t^2\right)$, which gives
\begin{equation}
    \tilde{\R}_2(S \gg 1)\approx \frac{L^2}{\sqrt{2\pi N}}\exp\left(-\frac{S^2}{2N}\right),
\label{eq:rho2_global}
\end{equation}
showing that the many-body energy spectrum has a width $w \sim \sqrt{N}$.

At the scale $\Delta\sim N^{-1}$ of single-particle level spacings, we must account for the contribution from $K_1$ (by setting $S\approx 0$ in Eq.~\eqref{eq:rho2_global}) as well as the ramp and plateau. We note that for $t$ near $2N$, we can expand the exponent in Eq.~\eqref{eq:K3}, obtaining, to leading order,
\begin{equation}
K_3(t < 2N) \approx L\exp\left[-\frac{(2N-t)^\frac{3}{2}}{3\pi N^{\frac{1}{2}}}\right].
\label{eq:K3approx3/2}
\end{equation}
$K_3(t)$ is therefore comparable to $L$ only in a relatively small region of size $\sim N^{1/3} \ll 2N$. This shows that for $N\to\infty$ and $t = O(N)$, $K(t)$ is given by a step function from $0$ to $L$ at the plateau time $t_\ast=2N$, i.e. $K(t) \to L \Theta(t-2N)$. To compute $\tilde{\R}_2(S\sim N^{-1})$ for large but finite $N$, we can  approximate $K_3(t)$ by a step function in $0\lesssim t<2N$, with the width chosen to enclose the same area up to the $t$-axis as in Eq.~\eqref{eq:K3approx3/2},
\begin{equation}
    K_{\text{ramp}}(t) \approx L\Theta(t-2N\alpha),
    \label{eq:ramptheta}
\end{equation}
%where
%\begin{equation}
where $\alpha = 1-\frac{2}{3}\Gamma\left(\tfrac{2}{3}\right)\left(3\pi/N\right)^\frac{2}{3}+O(N^{-3/5})$, as determined by integrating Eq.~\eqref{eq:K3approx3/2}, with $\alpha\to 1$ as $N\to \infty$.

We show in Sec.III of the supplementary material~\cite{Sup} that the generalization of $K_3(t)$ to an arbitrary (non-Gaussian) unitary random matrix ensemble~\cite{refRMTMehta} is the following upper bound for the many-body SFF at large $t$ (comparable to the inverse of the scale of single-particle level spacings),
\begin{equation}
    K(t) \leq L \exp\left[\frac{\kappa(t)-N}{4}\right],
    \label{eq:SFFupperbound}
\end{equation}
where $\kappa(t)$ is the single-particle SFF with a plateau time $t'_\ast$. For $t>t'_\ast$, the inequality becomes an equality, and $K(t)$ also attains a plateau. Consequently, as $0 < \left(N-\kappa(t<t'_\ast)\right) \sim O(N)$ in the exponent, $K(t)$ can be approximated by a step function expression similar to Eq.\eqref{eq:ramptheta}, with $2N\alpha$ replaced by $\sim t'_\ast$. Therefore, the form of the local contribution to many-body level statistics from Eq.\eqref{eq:ramptheta} is not just specific to the system defined by Eq.~\eqref{eq:P_h}, but should apply generally to a typical system with single-particle chaos in the unitary WD class.

\begin{figure}[!t]
\includegraphics[scale=0.4]{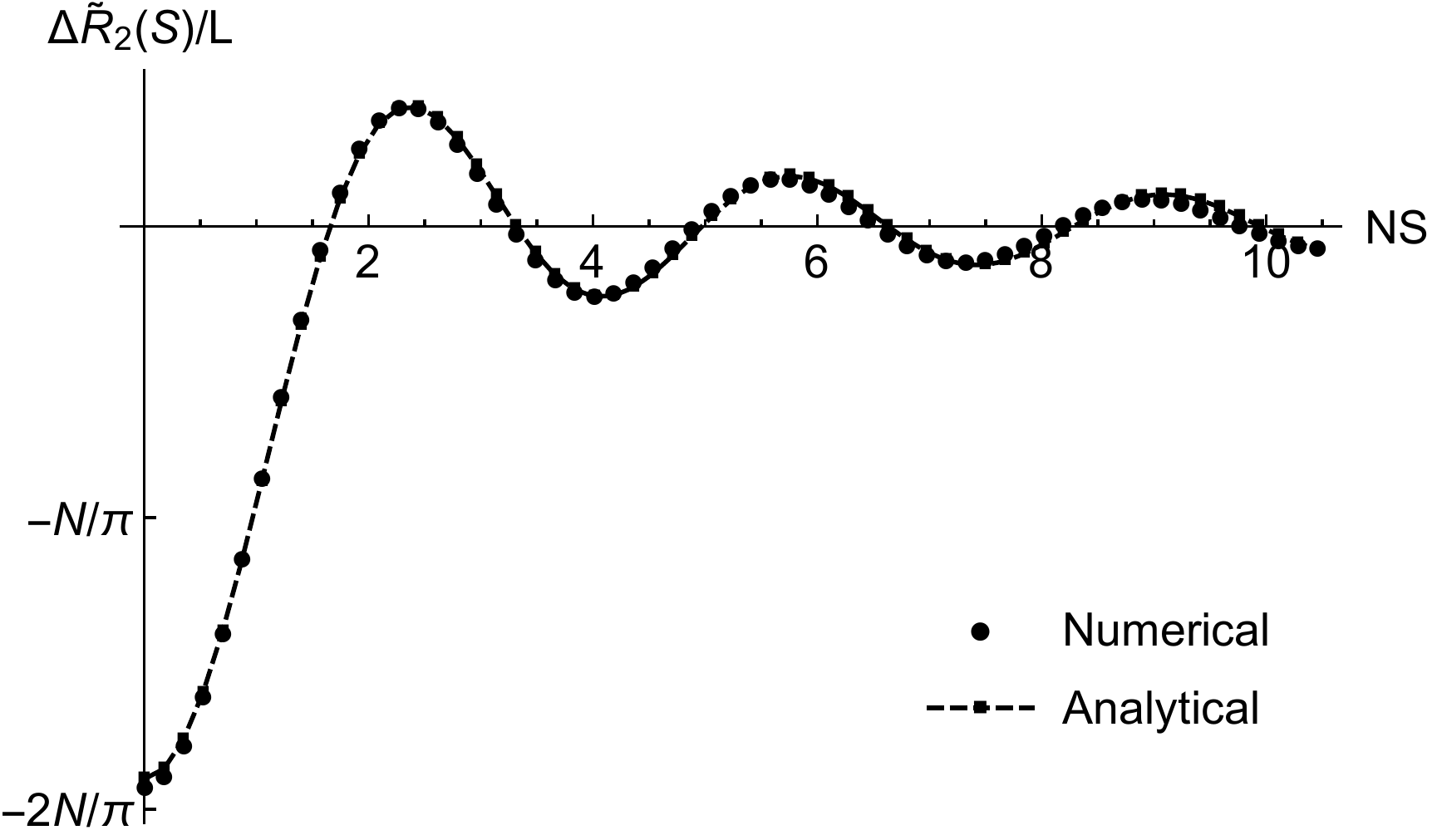}  
\caption{A plot of $\Delta\tilde{\R}_2(S)/L$ obtained from the Fourier transform of the SFF for $N=240$ as a function of scaled level separation $NS$, with the `analytical' curve from the second term of Eq.~\eqref{eq:MBlevrep} and the `numerical' data obtained from Eqs.~\eqref{eq:SFF_det},~\eqref{eq:wdef} and~\eqref{eq:rho2_def}.}
\label{fig:MBR2}
\end{figure}

Using Eq.~\eqref{eq:ramptheta} in Eq.~\eqref{eq:rho2_def}, together with a constant contribution at this scale from $K_1(t)$ gives
\begin{equation}
    \tilde{\R}_2(S \sim N^{-1}) \approx \frac{L^2}{\sqrt{2\pi N}}-\frac{2N\alpha L}{\pi}\frac{\sin(2N\alpha S)}{2N\alpha S}.
\label{eq:MBlevrep}
\end{equation}
The local contribution is the oscillatory second term, $\Delta\tilde{\R}_2(S) = \tilde{\R}_2(S)-L^2/\sqrt{2\pi N}$. This is plotted in Fig.~\eqref{fig:MBR2}, and compared with a numerical computation based on Eqs.~\eqref{eq:SFF_det} and~\eqref{eq:rho2_def}.
Equation~\eqref{eq:MBlevrep} can be contrasted with the more familiar two-level correlation function for the GUE (e.g.~\cite{refHaake, refRMTMehta, Cotler-2017, refRampPlateau2}),
\begin{equation}
    \R_2(S \sim N^{-1}) \propto 1-\frac{\sin^2(NS)}{N^2S^2}.
\label{eq:GUElevrep}
\end{equation}
The second term in both Eq.~\eqref{eq:MBlevrep} and Eq.~\eqref{eq:GUElevrep} contains the level repulsion effect, with any two levels least likely to have $S \ll N^{-1}$ at this scale. Unlike the single-particle GUE, where $\R_2(0) = 0$, the level repulsion for the many body case, Eq.~\eqref{eq:MBlevrep}, isn't total (i.e. $\tilde{\R}_2(0) \neq 0$), and in fact, negligible compared to the actual two-level density near $S=0$. This is essentially because the ultimate origin of this level repulsion is still the single particle level spectrum.

Another interesting feature of Eq.~\eqref{eq:MBlevrep} is that the second term can take both positive and negative values. This means that in addition to the level repulsion effect, there are less dominant centers of level attraction i.e. values of $S$ where levels tend to be found relative to each other with higher probability than in Poisson statistics (as determined by the asymptotic large-$S$ value within this regime). As $\alpha \to 1$ when $N\to \infty$, these essentially occur for values of $S$ immediately preceding the local maxima of Eq.~\eqref{eq:GUElevrep}.

In conclusion, we point out that an alternative to the brute-force algebraic (Hamiltonian) method to calculating the SFF, used here, is a Lagrangian formalism (a sigma model~\cite{Kamenev-GUE} or a similar path-integral approach to the SYK model~\cite{SYK-K,SYK-M}). As demonstrated in Supplement Sec.~IV~\cite{Sup} on the example of the sigma model, a non-perturbative resummation is required to recover the SFF. The zero mode fluctuation around the saddle point does detect the presence of an exponential ramp of Eq.~\eqref{eq:K2}~\cite{Brian}, but other soft modes and massive modes are equally important and contribute to the coefficient in the exponent. One can trace the presence of the sharp ramp in the non-interacting theory to infrared  divergences due to soft modes (the diffusons $(D{\bf q}^2 - i\omega)^{-1}$ in the theory of disordered metals, which reduce to $\omega^{-1}$ in zero-dimensional theories, such as random matrix theory studied here and equivalently SYK-2). In generic interacting theories, we expect the appearance of a dephasing-type cut-off in the soft modes that would result in suppression of the exponential ramp. This suppression is a necessary prerequisite for the expected transition to `true' many-body quantum chaos in the presence of interactions. As the Wigner-Dyson distribution would then be truly over the finely-spaced many body levels, the plateau time would be much closer to $t_\ast \sim L \gg N$, corresponding to a much slower, linear ramp in K(t). The details of the many-body SFF calculation for an interacting theory will be presented elsewhere.

\acknowledgements  
 Y.L. and V.G. acknowledge early conversations with Mark  Srednicki and  Chaitanya Murthy that motivated this project. V.G. is grateful to Sir Michael Berry for an illuminating discussion on many-body level statistics. The authors are grateful to   Brian Swingle, Michael Winer, and Shaokai Jian for many useful discussions and sharing their unpublished results.
This work was supported by the U.S. Department of Energy, Office of Science, Basic Energy Sciences under Award No. DE-SC0001911. Y. L. acknowledges a postdoctoral fellowship from the Simons Foundation “Ultra-Quantum Matter” Research Collaboration

\bibliography{main}

%\beginsupplement

\end{document}

\title{
Many-body level statistics of single-particle quantum chaos
\\
Supplemental Material}
\author{Yunxiang Liao}
\author{Amit Vikram}
\author{Victor Galitski}
\affiliation{Joint Quantum Institute, University of Maryland, College Park, MD 20742, USA.}
\affiliation{Condensed Matter Theory Center, Department of Physics, University of Maryland, College Park, MD 20742, USA}
\date{\today}
\maketitle

\maketitle

In this supplemental material, we show the detailed derivation of the spectral form factor (SFF) for an ensemble of noninteracting fermion systems described by the Hamiltonian
 \begin{align}\label{eq:H}
 \begin{aligned}
 	\hat{H}=\sum_{i,j=1}^{N} \hat{f}^{\dagger}_i (h_{ij}-\mu\delta_{ij}) \hat{f}_j,
 \end{aligned}
 \end{align}
 where $h$ is a random $N \times N$ Hermitian matrix from the Gaussian Unitary ensemble (GUE), which has the distribution function 
 \begin{align}\label{eq:P_h}
 \begin{aligned}
 	P(h) =2^{N(N-1)/2} \left( \frac{N}{2\pi}\right)^{N^2/2}  \exp \left[ -\frac{N}{2}\Tr \left( h^2 \right)  \right].
 \end{aligned}
 \end{align}
 The SFF $K(t)$ is defined as
 \begin{align}\label{eq:FT}
 \begin{aligned}
	 K (t) 
	 = \,&
	 \braket{Z(it)Z(-it)}
	 =
	 \braket{
	 	\sum_{\alpha,\beta}
	 	\exp \left[ -it \left( E_\alpha-E_\beta\right) \right] },
	 %\\
	 %=\,&
	 %\int dE_{\alpha} dE_{\beta}
	 %\R_2(E_\alpha, E_\beta)
	 %\exp \left[ -iT \left( E_\alpha-E_\beta \right) \right]
	 %+
	 %\textcolor{red}{L}
 \end{aligned}
 \end{align}
where $Z(it)=\Tr e^{ -i H t }$ is the partition function at imaginary inverse temperature $\beta =i t$, and the angular bracket stands for ensemble averaging.
 $\left\lbrace E_{\alpha}\right\rbrace $ represent the many-body energy levels which are related to the single-particle energy levels $\left\lbrace \e_i \right\rbrace $ through
 \begin{align}\label{eq:E}
	 E_{\alpha} =\sum_{n=1}^{N} n_i^{\alpha}\left( \e_i-\mu \right),
	 \qquad
	 n_i^{\alpha} = 0, 1.
 \end{align}	
Inserting Eq.~\eqref{eq:E} into the definition of the many-body SFF $K (t)$ Eq.~\eqref{eq:FT}, we find
\begin{align}\label{eq:FT-0}
\begin{aligned}
K (t)
=\,
\left\langle
		\sum_{\alpha,\beta}
		\exp 
		\left[ 
		-it \sum_{i=1}^{N} (\e_i-\mu) 
		\left( n_i^{\alpha}-n_i^{\beta}\right)  
		\right] 
\right\rangle
%	=\,
%\bigg\langle
%\prod_{i=1}^{N}
%\left\lbrace 
%\sum_{n_i^{\alpha/\beta}=0,1}
%\exp \left[ -iT  \e_i \left( n_i^{\alpha}-n_i^{\beta}\right)  \right] 
%\right\rbrace 
%\bigg\rangle
%\\
=\,
2^N
\bigg\langle
\prod_{i=1}^{N}
\left\lbrace 1+ \cos\left[ \left(\e_i-\mu\right) t  \right]  \right\rbrace 
\bigg\rangle,
\end{aligned}
\end{align}
which can be expressed as
\begin{align}\label{eq:FT-1}
\begin{aligned}
	&K (t)
	=\,
	2^N
	\int d \e_1...d \e_N
	P(\e_1,...,\e_N)
	\prod_{i=1}^{N}
	\left\lbrace 1+ \cos\left[ \left(\e_i-\mu\right) t  \right]  \right\rbrace .
\end{aligned}
\end{align}
Here $P(\e_1,...,\e_N)$ is the joint probability density function of the $N$ single-particle energy levels $\left\lbrace \e_i \right\rbrace $ and is given by
\begin{align}\label{eq:P_e}
	P(\e_1,\e_2,...,\e_N) 
	=
	C_N
	\exp (-\frac{N}{2} \e_i^2) \prod_{1\leq i<j \leq N} \lvert \e_i -\e_j \rvert^2,
%	\qquad
%	Z_N=\frac{N^{N(N-1)/4}}{(2\pi)^{N/2}} \prod_{i=1}^{N} \dfrac{\Gamma(2)}{\Gamma(1+j)}
\end{align}
where $C_N$ is a normalization constant.

The rest of the supplement is organized as follows. In Sec.~\ref{sec:I}, we discuss an approach based on the level correlation and cluster functions of the GUE, which lends itself well to estimating the initial slope region and the beginning of the ramp region (broadly, small-$t$ behavior). In Sec.~\ref{sec:II}, we instead express the SFF as a determinant involving Laguerre polynomials, which allows us to derive the ramp at very late times and the plateau time (broadly, large-$t$ behavior). In Sec.~\ref{sec:III}, we obtain an upper bound for the late-time many-body SFF in a general unitary ensemble and derive its plateau time, which is used to argue for the generality of the many-body level statistics derived in the main text. In Sec.~\ref{sec:IV}, we consider a path integral based $\sigma$-model, and examine how much of the behavior of the SFF can be extracted from this method - which has the advantage that it has a more natural generalization to systems of interacting fermions.

\section{Cluster function approach}\label{sec:I}

\subsection{A.\ Correlation and Cluster Functions }

Using the fact that the joint probability density is symmetric under the permutation of its arguments, i.e., $P(..,\e_i,..,\e_j,...)=P(..,\e_j,..,\e_i,...)$, we rewrite Eq.~\eqref{eq:FT-1} as
\begin{align}\label{eq:FT1}
\begin{aligned}
	&K (t)
	=\,
%	2^N
%	\int \prod_{i=1}^{N} d \e_i
%	P(\e_1,...\e_N)
%	\left[
%	1+
%	\sum_{n=1}^{N}
%	\frac{N!}{(N-n)!n!}
%	\prod_{i=1}^{n}\cos\left( \e_i T\right) 
%	\right] 
%	\\
%	=\,&
	2^N
	\left[ 1
	+
	\sum_{n=1}^{N}
	\frac{1}{n!}
	\int
	\R_n(\e_1,...,\e_n)
	\prod_{i=1}^{n}
	\cos\left[ \left(\e_i-\mu\right) t  \right] 
	 d \e_i 
	\right].
\end{aligned}
\end{align}
Here $\R_n(\e_1,...,\e_n)$ is the $n$-point single-particle energy level correlation function, and is defined as
\begin{align}\label{eq:Rn}
\R_n(\e_1, ....,\e_n)
=\,
\frac{N!}{(N-n)!}\int d\e_{n+1} ...d\e_N  P(\e_1, ..., \e_N).
\end{align}
It gives the probability density of finding a energy level around each $\e_i$, for $i=1,..n$, irrespective of the remaining levels and independent of the labeling.
It has been found that, for energy level described by the GUE probability distribution (Eq.~\eqref{eq:P_e}), 
%the $n$-point single particle energy level correlation function 
$\R_n(\e_1, \e_2....\e_n)$ is given by the determinant of the kernel $\kk(\e_i,\e_j)$~\cite{refRMTMehta},
\begin{align}\label{eq:Rn}
\begin{aligned}
\R_n(\e_1,\e_2...\e_n)
=\,
\det \left[  \kk(\e_i,\e_j) \right]_{i,j=1,...n},
\end{aligned}
\end{align}
where $\kk(\e_i,\e_j)$, in the large $N$ limit, takes the form of
\begin{align}\label{eq:KK}
\begin{aligned}
	\kk(\e_i,\e_j)
	=
	\begin{cases}
	\R_1(\e_i)
	=
	\dfrac{N}{2\pi} \sqrt{4-\e_i^2} \, \Theta (2-|\e_i|),
	&
	i=j.
	\\
	\\
	\kk(\e_i-\e_j)=\dfrac{N}{\pi}\dfrac{\sin \left[ N\left(\e_i-\e_j \right)  \right] }{N \left(\e_i-\e_j \right)},
	&
	i \neq j.
\end{cases}
\end{aligned}
\end{align}
Here $\R_1(\e_i)$ is the average single-particle level density and exhibits Wigner's semicircle law.
We note that the $2$-point many-body SFF $K(t)$ is given by the summation of the Fourier transform of the $n$-point single-particle energy level correlation function which is closely related to the single-particle $n$-point SFF, with $n$ running over $1,...N$.
%We note that
%\begin{align}\label{eq:cos}
%\begin{aligned}
%	\prod_{i=1}^{n}\cos\left( \e_i T\right) 
%	=\frac{1}{2^n} \sum_{\left\lbrace \zeta_i=\pm 1 \right\rbrace } \exp \left(i  t \sum_{i=1}^{n} \e_i \zeta_i \right),
%\end{aligned}
%\end{align}
%and as a result, the many-body $2$-point SFF $K(t)$ is given by the summation of the single-particle $n$-point SFF, with $n$ running over $n=1,...N$. $\suml{\left\lbrace \zeta_i=\pm 1 \right\rbrace }$ sums over all possible configuration with $\zeta_i=1$ or $-1$.

%\subsection{Cluster functions}

It is convenient to define the $n$-point cluster function $\Tm_n(\e_1,...,\e_n)$ by excluding the lower order correlation~\cite{refRMTMehta}:
\begin{align}\label{eq:cluster}
\begin{aligned}
	\Tm_n(\e_1,...,\e_n)
	=&
	\suml{ \GG  } (-1)^{n-|\GG|} (|\GG|-1)! 
%	\\
%	\times&
	\prod_{j=1}^{|\GG|} \R_{|\GG_j|}(\left\lbrace \e_k,k \in \GG_j\right\rbrace ).
\end{aligned}
\end{align}
Here $\GG$ represents a partition of the indices $\left\lbrace 1,2,...,n\right\rbrace $ into $|\GG|$ subgroups $\left\lbrace \GG_i, i=1,...|\GG| \right\rbrace $, with each group $\GG_i$ of length $|\GG_i|$. It obeys the constraint $\sum_{i=1}^{|\GG|}|\GG_i|=n$, with $|\GG|$ being the number of subgroups in the partition.
From Eqs.~\eqref{eq:cluster} and~\eqref{eq:Rn}, one can deduce the form of the  $n$-point cluster-function $\Tm_n$~\cite{refRMTMehta}:
\begin{align}\label{eq:Tn}
\begin{aligned}
\Tm_n(\e_1,...\e_n)
=\,
\suml{\PP(n)} \kk(\e_{1},\e_{2})\kk(\e_{2},\e_{3})...\kk(\e_{n-1},\e_{n})\kk(\e_{n},\e_{1}),
\end{aligned}
\end{align}
where the summation is over $(n-1)!$ cyclic permutations $\PP(n)$ of indices $\left\lbrace 1,2,...n\right\rbrace $.

We then define
\begin{align}\label{eq:rtn}
\begin{aligned}
	&\bar{\rr}_n
	=	\int  d \e_1 d \e_2 ...d \e_n
	\R_n(\e_1,\e_2,...,\e_n)
	\prod_{i=1}^{n} \cos\left[ \left(\e_i-\mu\right) t  \right]  ,
	\\
	&\bar{\tm}_n
	=	\int  d \e_1 d \e_2 ...d \e_n
	\Tm_n(\e_1,\e_2,...,\e_n)
	\prod_{i=1}^{n} \cos\left[ \left(\e_i-\mu\right) t  \right] .
\end{aligned}
\end{align}
The many-body SFF $K(t)$ can now be expressed as
\begin{align}\label{eq:FT-2}
\begin{aligned}
	K (t)
	=2^N \left[  1+\sum_{n=1}^{N} \frac{1}{n!} \bar{\rr}_n\right] 
	=2^N \exp \left[ \sum_{n=1}^{N} (-1)^{n-1} \frac{1}{n!}\bar{\tm}_n \right],
\end{aligned}
\end{align}
where in the second equality, we have approximated the upper limit $N$ with infinity, and use the relation between the correlation function $R_n(\e_1,...,\e_n)$ and the cluster function $T_n(\e_1,...,\e_n)$ (Eq.~\eqref{eq:cluster}).
%\subsection{Many-body SFF}
		
%We then combine Eqs~\eqref{eq:Tn} and~\eqref{eq:rtn} to calculate $\bar{\tm}_n$, which is then inserted into Eq.~\eqref{eq:FT-2} to obtain the SFF $K(t)$. 

\subsection{B. Calculation of $\tm_n$}	\label{sec:tm}

In this section, we calculate $\bar{\tm}_n$ defined in Eq.~\eqref{eq:rtn}, which can be used to obtain the SFF $K(t)$ using Eq.~\eqref{eq:FT-2}.
Substituting Eq.~\eqref{eq:Tn} into Eq.~\eqref{eq:rtn} leads to
\begin{align}\label{eq:tn}
	\begin{aligned}
		\bar{\tm}_n
		=\,&
		\frac{1}{2^n} \sum_{\left\lbrace \zeta_i=\pm 1 \right\rbrace} 
		\int  d \e_1 d \e_2 ...d \e_n e^{i  t \sum_{i=1}^{n}(\e_i -\mu)\zeta_i }
%		\\
%		\times &
		\left[ 
		\suml{\PP(n)}
		\kk(\e_{1},\e_{2})\kk(\e_{2},\e_{3})...\kk(\e_{n-1},\e_{n})\kk(\e_{n},\e_{1})
		\right] 	
		\\
		=\,&
		\frac{(n-1)!}{2^n} \sum_{\left\lbrace \zeta_i=\pm 1, i=1,..,n \right\rbrace } 
		I_n(\left\lbrace \zeta_i \right\rbrace ) e^{-i  t\mu \sum_{i=1}^{n}\zeta_i },
	\end{aligned}	
\end{align}	
where $I_n(\left\lbrace \zeta_i \right\rbrace )$ is defined as
\begin{align}\label{eq:In-0}
	\begin{aligned}
		&I_n(\left\lbrace \zeta_i \right\rbrace )
		\equiv 
		\int  d \e_1 ...d \e_n
		\kk(\e_{1},\e_{2})\kk(\e_{2},\e_{3})...\kk(\e_{n},\e_{1})
%		\\
%		&\times 
		e^{i  t \sum_{i=1}^{n} \e_i \zeta_i }.
	\end{aligned}	
\end{align}

For $n=1$, it is easy to see that $I_1(\zeta_1)$ is the Fourier transform of the average single-particle level density $\R_1(\e)$:
\begin{align}\label{eq:I1}
	I_1(\zeta_1)
	=\,&
	\int d \e \, \R_1(\e)  \exp \left(i  t  \e \zeta_1 \right) 
	=
	N \dfrac{J_1(2t)}{t},
\end{align}
where $J_1(x)$ is the Bessel function of the first kind and admits the asymptotic form of $J_1(x)=\sqrt{2/\pi x} \cos (x-3\pi/4)$ for $x\rightarrow \infty$.	

To evaluate $I_n(\left\lbrace \zeta_i \right\rbrace )$ for $n \geq 2$, we perform the following transformation
\begin{align}
	\begin{aligned}
		u_i
		=
		\begin{cases}
			\e_{i}-\e_{i+1},
			&
			i=1,...,n-1,
			\\
			\e_n,
			&
			i=n.
		\end{cases}
	\end{aligned}
\end{align}
Under this transformation, one has
\begin{align}
	\sum_{i=1}^{n} \e_i \zeta_i
	=
	\sum_{i=1}^{n} \left( \sum_{j=i}^{n} u_j \right) \zeta_i 
	=
	\sum_{i=1}^{n}  u_i \left( \sum_{j=1}^{i} \zeta_j\right) ,
\end{align}
which leads to
\begin{align}\label{eq:FTK-1}
	\begin{aligned}
		&I_n(\left\lbrace \zeta_i \right\rbrace )
		=\, 
		\int_{-\infty}^{\infty}..\int_{-\infty}^{\infty} 	d u_1...d u_{n-1} \int_{-\pi/2}^{\pi/2} d u_n
%		\\
%		\times &
		\kk(u_1)...\kk(u_{n-1})\kk(-\sum_{i=1}^{n-1}u_i)
		e^{ i  t \sum_{i=1}^{n}  u_i \left( \sum_{j=1}^{i} \zeta_j\right)  }.
	\end{aligned}
\end{align}
Here we have employed the box approximation explained in Ref.~\cite{refRampPlateau,refRampPlateau2}. 
We note that $K(\e_i \neq \e_j)$ takes the form of Eq.~\eqref{eq:KK} only for $\e_{i,j}$ close to the origin $\e=0$ such that $\R_1(\e)$ can be approximated by $\R_1(0)$. As a result, we have to impose a cut-off $|\e| \leq \pi/2$ which is determined by the normalization condition
$\int_{-\pi/2}^{\pi/2}\R_1(0)=N$. 
We then extend the integration region of $u_j, j=1,...n-1$ to the entire space due to the presence of $K(u_j)$ which decays rapidly as $|u_j|$ increases.

After rewriting $\kk(-\sum_{i=1}^{n-1}u_i)$ as a two variable integral
\begin{align}
	\begin{aligned}
		&\kk(-\sum_{i=1}^{n-1}u_i)
		=
		\int_{-\infty}^{\infty} d u \kk(u)\delta(u+\sum_{i=1}^{n-1}u_i)
		=
		\int_{-\infty}^{\infty} du  \kk(u) \int_{-\infty}^{\infty} \frac{dk}{2\pi} \exp \left[i k \left( u+\sum_{i=1}^{n-1}u_i \right)  \right].
	\end{aligned}
\end{align}
and inserting it into Eq.~\eqref{eq:FTK-1}, one arrives at
\begin{widetext}
	\begin{align}\label{eq:FTK-2}
		\begin{aligned}
			&I_n(\left\lbrace \zeta_i \right\rbrace )
			=\,
			\int_{-\pi/2}^{\pi/2} d u_n e^{i  u_n t \sum_{j=1}^{n} \zeta_j }
			\int_{-\infty}^{\infty} \frac{dk}{2\pi} 
			\int_{-\infty}^{\infty}..\int_{-\infty}^{\infty}  d u_1...d u_{n-1} du  
			\,
			e^{iku+ i   \sum_{i=1}^{n-1}  u_i \left( t \sum_{j=1}^{i} \zeta_j+k\right) }
			\kk(u) 
			\prod_{i=1}^{n-1}\kk(u_i) 	
			\\
			=&
			\pi \dfrac{\sin \left[  \frac{\pi}{2} \left( t \sum_{j=1}^{n} \zeta_j \right) \right]  }
			{ \frac{\pi}{2} \left( t \sum_{j=1}^{n} \zeta_j \right) }
			\int_{-\infty}^{\infty} \frac{dk}{2\pi} \,
			\tilde{\kk}(k) 
			\prod_{j=1}^{n-1}
			\tilde{\kk}(k+t\sum_{i=1}^{j}\zeta_i),
		\end{aligned}
	\end{align}
\end{widetext}	
where $\tilde{\kk}(k)$ denotes the Fourier transform of the kernel $\kk (\e) $
\begin{align}\label{eq:tK}
	\begin{aligned}
		\tilde{\kk}(k) \equiv \int_{-\infty}^{\infty} d u\, \kk (u) e^{iuk}
		=&
		\Theta \left( 1-\bigg\lvert \frac{k}{N} \bigg\rvert \right).
	\end{aligned}
\end{align}
Substituting the explicit form of $\tilde{\kk}(k)$ (Eq.~\eqref{eq:tK}) into Eq.~\eqref{eq:FTK-2} then yields the following expression for $I_n(\left\lbrace \zeta_i \right\rbrace )$ when $n \geq 2$:
\begin{align}
\begin{aligned}\label{eq:I-n}
I_{n\geq 2}( \left\lbrace \zeta_i \right\rbrace )
=&
N
\dfrac{\sin \left[  \frac{\pi}{2} \left( t \sum_{j=1}^{n} \zeta_j \right) \right]  }
{ \frac{\pi}{2} \left( t \sum_{j=1}^{n} \zeta_j \right) }
\left[    1-\dfrac{t}{2N} s(\left\lbrace \zeta_i\right\rbrace ) \right] 
%\\
%\times &
\Theta 	\left[   1-\dfrac{t}{2N} s(\left\lbrace \zeta_i\right\rbrace )\right].
\end{aligned}
\end{align}
Here $s(\left\lbrace \zeta_i \right\rbrace )$ is defined as the difference between the maximum and the minimum of the series $\left\lbrace 0,\sum_{i=1}^{j} \zeta_i\right\rbrace$, for $j$ running over  $j=1,...n-1$:
	\begin{align}\label{eq:s}
	\begin{aligned}
	s(\left\lbrace \zeta_i \right\rbrace )
	=
	\max\left\lbrace 0,\sum_{i=1}^{j}\zeta_i \right\rbrace_{j=1}^{n-1}
%	\\
	-
	\min \left\lbrace 0,\sum_{i=1}^{j} \zeta_i \right\rbrace_{j=1}^{n-1}.
	\end{aligned}
	\end{align}

\subsection{C. General expression for the SFF}

From previous calculation, we find the SFF $K(t)$ can be expressed as
\begin{align}\label{eq:FT-3}
\begin{aligned}
	&K (t)
	=\,
	2^N \exp \left[ \sum_{n=1}^{N} (-1)^{n-1} \frac{1}{n}\frac{1}{2^n} \sum_{\left\lbrace \zeta_i=\pm 1\right\rbrace } 
	I_n(\left\lbrace \zeta_i \right\rbrace ) 
	e^{-i  t\mu \sum_{i=1}^{n}\zeta_i }
	\right],
\end{aligned}
\end{align}
where $\suml{\left\lbrace \zeta_i=\pm 1 \right\rbrace}$ denotes the summation over all possible configurations with $\zeta_i$ taking the value of $+1$ or $-1$ for $i=1,..,n$.
 $I_n ( \left\lbrace \zeta_i \right\rbrace ) $ is given by Eq.~\eqref{eq:I1} and Eq.~\eqref{eq:I-n} for $n=1$ and $n \geq 2$, respectively.
%	\begin{subequations}
%	\begin{align}
%		&\label{eq:I-1}
%		I_1(\zeta_1)
%		=
%		N \dfrac{J_1(2t)}{t},
%		\\
%		&
%		\begin{aligned}\label{eq:I- }
%		I_{n\geq 2}( \left\lbrace \zeta_i \right\rbrace )
%		=&
%		N
%		\dfrac{\sin \left[  \frac{\pi}{2} \left( t \sum_{j=1}^{n} \zeta_j \right) \right]  }
%		{ \frac{\pi}{2} \left( t \sum_{j=1}^{n} \zeta_j \right) }
%		 \left[    1-\dfrac{t}{2N} s(\left\lbrace \zeta_i\right\rbrace ) \right] 
%		 \\
%		 \times &
%		\Theta 	\left[   1-\dfrac{t}{2N} s(\left\lbrace \zeta_i\right\rbrace )\right].
%		\end{aligned}
%	\end{align}
%	\end{subequations}

	At time $t=0$ and $t \rightarrow \infty$, $I_n(\left\lbrace \zeta_i \right\rbrace )$ takes the value of $N$ and $0$, respectively, for arbitrary configuration $\left\lbrace \zeta_i \right\rbrace$, which leads to $K(0)=L^2$ and $K ( \infty)=L$, as expected from the definition of the SFF (Eq.~\eqref{eq:FT}).
	At large time $t \gg N$, we find that $I_n(\left\lbrace \zeta_i\right\rbrace)=0$ for $n \geq 2$ due to presence of the unit step function, which yields
	\begin{align}
	\begin{aligned}
		K(t \gg N)
		=
		2^N 
		\exp 
		\left( 
		N \dfrac{J_1(2t)}{t} \cos (\mu t)
		\right)
		\approx 
		L.
	\end{aligned}
	\end{align}

	To further simplify the calculation of $K(t)$, we then group together $I_n(\left\lbrace \zeta_i\right\rbrace )$ with the same value of $p=\sum_{i=1}^{n} \zeta_i$, and rewrite $K(t)$ as
	\begin{widetext}
	\begin{align}\label{eq:FT-4}
	\begin{aligned}
	&K (t)
	=\,
	2^N 
	\exp 
	\left\lbrace 
	N \dfrac{J_1(2t)}{t}
	\cos ( \mu t)
	+
	A_0(t)
	+
	2
	\sum_{p=1}^{N}
	A_p(t)  (-1)^p
	\dfrac{\sin \left(  \pi t p/2  \right)   }{ \pi t p/2  }
	\cos (p \mu t)
	\right\rbrace ,
	\end{aligned}
	\end{align}
	where $A_p(t)$ is defined as
	\begin{align}\label{eq:Ap}
	\begin{aligned}
	A_p(t)
	\equiv
	-
	N
	\sum_{n=2}^{N}  \frac{1}{n} \frac{1}{2^n} 
	\sum_{\sum_{i=1}^{n} \zeta_i =p} 
	\left[ 1-\dfrac{t}{2N}   s(\left\lbrace \zeta_i\right\rbrace )  \right] 
	\Theta 	\left[   1 - \dfrac{t}{2N}  s(\left\lbrace \zeta_i\right\rbrace) \right].
	\end{aligned}
	\end{align}	
	\end{widetext}
	$\suml{\sum_{i=1}^{n} \zeta_i =p} $ represents the summation over all configurations of $\left\lbrace \zeta_i=\pm 1\right\rbrace_{i=1}^{n}$ obeying the constraint $\sum_{i=1}^{n} \zeta_i =p$.
	We have used the fact that $\sum_{i=1}^{n} \zeta_i=p$ can only be satisfied for even $n-p$, and also $A_p(t)=A_{-p}(t)$ due to symmetry.

	Eq.~\eqref{eq:FT-4} is valid at all time $t$. We now turn to the regime $t \ll N$, and approximate all unit step functions in $A_p(t)$ with $1$.
	We note that for a small number of configurations, $s(\left\lbrace \zeta_i\right\rbrace )$ is of the order of $N$, and $\Theta  \left[   1-\dfrac{t}{2N}  s(\left\lbrace \zeta_i\right\rbrace) \right]$ becomes zero for $t \gg 1$. However, because of the overall factor $1/n2^{n}$ in the summation and the fact that the number of configurations with $s(\left\lbrace \zeta_i\right\rbrace )\sim O(N)$ is small, we ignore such situations, and set all $\Theta \left[   1-\dfrac{t}{2N}  s(\left\lbrace \zeta_i\right\rbrace) \right]$ in Eq.~\eqref{eq:Ap} to $1$.
	$A_p(t)$ takes the form of
	\begin{subequations}
	\begin{align}\label{eq:Ap2}
	&A_p(t)
	=
	-B_p+C_p t,
	\\
	&\label{eq:Bp}
	B_p
	\equiv
	N
	\sum_{m=m_0}^{N} 
	\frac{1}{2m+|p|} \frac{1}{2^{2m+|p|}} 
	\frac{(2m+|p|)!}{(m+|p|)!m!},
	\\
	&\label{eq:Cp}
	C_p
	\equiv
	\sum_{n=2}^{N}
	\frac{1}{n} \frac{1}{2^{n+1}} 
	\sum_{\sum_{i=1}^{n} \zeta_i =p} 	
	 s(\left\lbrace \zeta_i\right\rbrace ),
	\end{align}	
	\end{subequations}
	where the lower limit $m_0$ in the summation of Eq.~\eqref{eq:Bp}  is given by $0$ for integer $|p| >1$, and $1$ for $|p |\leq 1$.
	
It is straightfoward to see that, in the large N limit, $B_p$ is given by
	\begin{align}\label{eq:Bp-1}
	\begin{aligned}
	B_p
	=
	\begin{dcases}
	N\ln 2,
	&
	p=0,
	\\
	\frac{N}{2},
	& 
	|p|=1,
	\\
	\frac{N}{|p|},
	&
	|p| >1.
	\end{dcases}	
	\end{aligned}
	\end{align}
By contrast, the calculation of $C_p$(Eq.~\eqref{eq:Cp}) is complicated and can be mapped to a random walk problem. We rewrite the summation of $s(\left\lbrace \zeta_i \right\rbrace)$ as the difference of two separate summations:
\begin{align}\label{eq:sums}
	\begin{aligned}
		\sum_{\sum_{i=1}^{n} \zeta_i=p} s(\left\lbrace \zeta_i \right\rbrace)
		=
		\sum_{\sum_{i=1}^{n} \zeta_i=p} s_a(\left\lbrace \zeta_i \right\rbrace)
		-
		\sum_{\sum_{i=1}^{n} \zeta_i=p} s_b(\left\lbrace \zeta_i \right\rbrace),
	\end{aligned}
\end{align}
where
\begin{align}\label{eq:sab}
	\begin{aligned}
		s_a(\left\lbrace \zeta_i \right\rbrace )
		=\max\left\lbrace 0, \sum_{i=1}^{j} \zeta_i \right\rbrace_{j=1}^{n-1}, 
		\qquad
		s_b(\left\lbrace \zeta_i \right\rbrace )
		=\min \left\lbrace 0, \sum_{i=1}^{j} \zeta_i\right\rbrace_{j=1}^{n-1}.
	\end{aligned}
\end{align}

\subsection{D. Calculation of $C_0$}

Let us first consider the case $p=0$. 
Because of the symmetry, the two sums on the right-hand side of Eq.~\eqref{eq:sums} are opposite with respect to each other and their difference
can be further reduced to 
\begin{align}
	\begin{aligned}
		\sum_{\sum_{i=1}^{n} \zeta_i=0} s(\left\lbrace \zeta_i \right\rbrace)
		=
		2\sum_{\sum_{i=1}^{n} \zeta_i=0} s_a(\left\lbrace \zeta_i \right\rbrace),
	\end{aligned}
\end{align}
where $n$ has to be an even integer.

To evaluate this summation, for each configuration of $\left\lbrace \zeta_i \right\rbrace_{i=1}^{n}$, we introduce a series $\left\lbrace x_i \right\rbrace $:
\begin{align}\label{eq:x}
	\begin{aligned}
		x_i
		=\begin{cases}
			0, & i=0,
			\\
			\sum_{j=1}^{i} \zeta_j, & i=1,...n.
		\end{cases}
	\end{aligned}
\end{align}
where $x_i$ represents the position at step $i$.
After each step, $x_{i}$ is increased by $\zeta_{i+1}$ which can only takes the value of $+1$ or $-1$: $x_{i+1}=x_{i}+\zeta_{i+1}$.
Each configuration of $\left\lbrace \zeta_i \right\rbrace $ satisfying the constraint $\sum_{i=1}^{n} \zeta_i=0$ corresponds to a path that starts from $x_0=0$ and ends at $x_{n}=0$ after $n$ steps. 
The number of such paths is
$(n)!/(n/2)!(n/2)!$.
In addition, $s_a(\left\lbrace \zeta_i \right\rbrace )=\max\left\lbrace x_i \right\rbrace_{i=0}^{n-1} $ is the maximum position reached before the last step for path $\left\lbrace x_i \right\rbrace $. 

\begin{figure}[t!]
	\centering
	\includegraphics[width=0.4\linewidth]{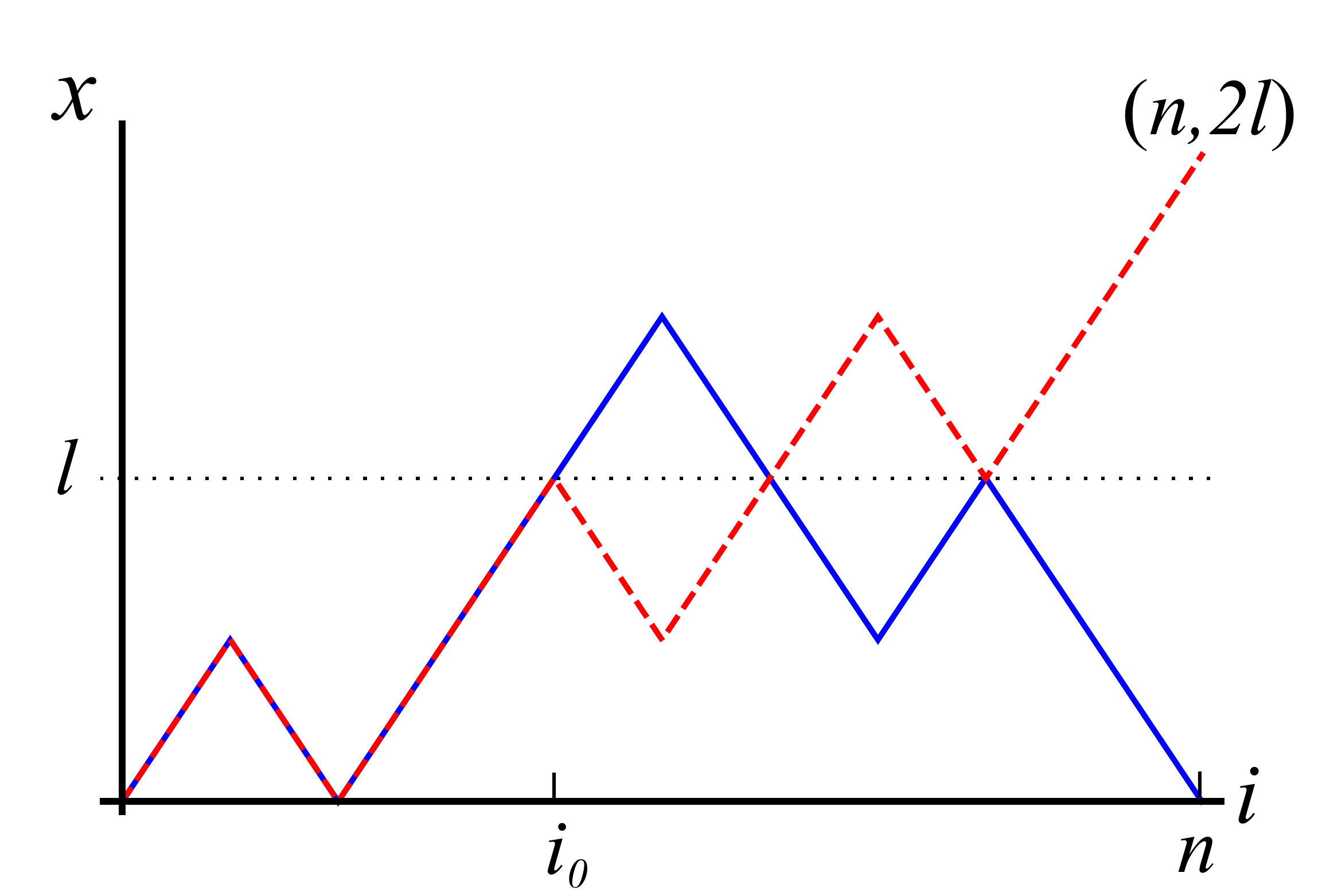}
	\caption{Reflection path. The blue solid lines correspond to the original path defined by Eq.~\eqref{eq:x}, which begins from $(i=0,x_0=0)$ and ends at $(i=n,x_n=0)$. The red dashed lines represent the corresponding reflected path defined by Eq.~\eqref{eq:reflect}, which ends at $(i=n,x_n=2l)$ instead. Starting from $i_0$ the first step at which the original path reaches $x_{i_0}=l$, the original path (blue solid lines) is reflected around the line $x=l$ (gray dashed line) to obtain the reflected path (red dashed lines).
	}
	\label{fig:walk}
\end{figure}

In order to calculate the summation of $s_a(\left\lbrace \zeta_i \right\rbrace )$ for all the paths that start from $x_0=0$ and end at $x_{n}=0$, we count the total number of the paths for which $s_a(\left\lbrace \zeta_i \right\rbrace ) \geq l$, denoted as $N_n^{(0)}(s_a(\left\lbrace \zeta_i \right\rbrace ) \geq l)$.
If we plot $x_i$ as a function of the step $i$ for all the paths (see Fig.~\eqref{fig:walk}),
those with $s_a(\left\lbrace \zeta_i \right\rbrace )\geq l$ should cross the line $x=l$ at least once at some intermediate steps, the smallest among which is called $i_0$.
For each of these paths $\left\lbrace x_i \right\rbrace $, we now define its reflection path $\left\lbrace x'_i \right\rbrace $
where every step after $i > i_0$ is reversed. The reflection path is symmetric around $x=l$ for $i >i_0$, while for $i \leq  i_0$, it remains the same:
\begin{align}\label{eq:reflect}
	\begin{aligned}
		x_i'
		=
		\begin{cases}
			x_i,  & i \leq i_0,
			\\
			2l-x_i & i > i_0.
		\end{cases}
	\end{aligned}
\end{align}
As shown in Fig.~\eqref{fig:walk},
the reflection path (represented by the red dashed line) then reach $x_{n}'=2l$ after $n$ steps since the original path (represented by the blue solid line) ends at $x_{n}=0$.
The total number of reflection paths that reach $x_{n}'=2l$ 
is the same to the number of the original paths that cross the line $x=l$ (which starts at $x_0=0$ and ends at $x_{n}=0$), and equals the number of the paths with $s_a(\left\lbrace \zeta_i \right\rbrace )\geq l$:
\begin{align}
	N_n^{(0)}(s_a(\left\lbrace \zeta_i \right\rbrace ) \geq l)=\frac{n!}{(\frac{n}{2}+l)!(\frac{n}{2}-l)!}.
\end{align}
Here $l$ is a positive integer with $1 \leq l \leq n/2$ .

We note that
\begin{align}\label{eq:sumN}
	\begin{aligned}	
		&\sum_{l=1}^{n/2}N_n^{(0)}(s_a(\left\lbrace \zeta_i \right\rbrace ) \geq l)
		=\sum_{l=1}^{n/2}\sum_{k=l}^{n/2} N_n^{(0)}(s_a(\left\lbrace \zeta_i \right\rbrace ) = k)
%		\\
		=
		\sum_{k=1}^{n/2} k N_n^{(0)}(s_a(\left\lbrace \zeta_i \right\rbrace ) = k),
	\end{aligned}	
\end{align}
where $N_n^{(0)}(s_a(\left\lbrace \zeta_i \right\rbrace ) = k)$ denotes the number of path with $s_a(\left\lbrace \zeta_i \right\rbrace ) = k$.
The above equation gives
\begin{align}\label{eq:Gsum}
	\begin{aligned}
		\sum_{\sum_{i=1}^{n} \zeta_i=0} 
		s_a(\left\lbrace \zeta_i\right\rbrace )
		=
		\sum_{l=1}^{n/2} 	\frac{n!}{(\frac{n}{2}+l)!(\frac{n}{2}-l)!}
%		\\
		=
		\frac{1}{2}
		\left( 
		2^{n}-\frac{n!}{(\frac{n}{2})!(\frac{n}{2})!}
		\right).
	\end{aligned}
\end{align}

Inserting Eq.~\eqref{eq:Gsum} into Eq.~\eqref{eq:Cp} while setting $p=0$, one obtains, in the large $N$ limit,
\begin{align}\label{eq:C0}
	\begin{aligned}
		C_0
		=&
		\sum_{m=1}^{\floor{N/2}} \frac{1}{4m} \frac{1}{2^{2m}} 
		\left[ 2^{2m}-\frac{(2m)!}{m!m!} \right] 
		=
		\sum_{m=1}^{\floor{N/2}} \frac{1}{4m}
		-
		\frac{1}{2}\ln 2
		\\
		=\,&
		\frac{1}{4}
		\left[ 
		\psi_0(N/2+1)
		+\gamma_E
		\right]  
		-
		\frac{1}{2}
		\ln 2
		=\frac{1}{4}
		\left( 
		\ln \frac{N}{8} +\gamma_E
		\right),
	\end{aligned}
\end{align}
where $\psi_0(x)$ is the digamma function and $\gamma_E \approx 0.577$ is the Euler-Mascheroni constant.

%\begin{align}\label{eq:A_0}
%\begin{aligned}
%A_0(T \ll N)
%=
%-N\ln 2
%+
%\frac{1}{4}
%\left( 
%\ln \frac{N}{8} +\gamma_E
%\right) 
%T.
%\end{aligned}
%\end{align}

\subsection{E. Calculation of $C_p$ for $p>0$}\label{sec:Cp}

$C_p$ for $p>0$ can be evaluated in an analogous way as $C_0$. 
We now consider the paths $\left\lbrace x_i \right\rbrace $ (defined in Eq.~\eqref{eq:x}) that start from $x_0=0$ and end at $x_n=p$ (therefore satisfy $\sum_{i=1}^{n} \zeta_i=p$), for non-negative even $n-p$.
$s_a(\left\lbrace \zeta_i \right\rbrace )=\max\left\lbrace x_i\right\rbrace_{i=0}^{n-1} $ is again the maximum position reached before the $n$-th step.

Using the reflection path method introduced in the previous section,
we calculate the number of such paths with $s_a(\left\lbrace \zeta_i \right\rbrace ) \geq l$, denoted as $N_n^{(p)}(s_a(\left\lbrace \zeta_i \right\rbrace )\geq l)$. 
In the $i-x$ plane, these paths cross the line $x=l$ at least once for $ p \leq l \leq \frac{n+p}{2}$.  Their reflection paths defined by Eq.~\eqref{eq:reflect} should reach $x_n=2l-p$ after $n$ steps, and can be used to find the total number of path with $s_a(\left\lbrace \zeta_i \right\rbrace ) \geq l$ for $p+1 \leq l \leq \frac{n+p}{2}$:
\begin{align}\label{eq:Nnp}
	N_n^{(p)}(s_a(\left\lbrace \zeta_i \right\rbrace )\geq l)
	=
	\dfrac{n!}{(\frac{n-p}{2}+l)!(\frac{n+p}{2}-l)!}.
\end{align}

For $l=p$, the above expression is no longer valid. 
Instead, setting $l$ on the right-hand side of the above equation to $p$ gives $N_n^{(p)}(s_a(\left\lbrace \zeta_i \right\rbrace ) \geq p-1)$, which is also the total number of $n$-step paths with $x_0=0$ and $x_{n}=p$.
We have to consider the situation $s_a(\left\lbrace \zeta_i \right\rbrace )= p$ separately. In this case, it is easy to see that $x_{n-1}=p-1$. The number of paths that begin from $x_0=0$ and end at $x_n=p$ while satisfying $s_a(\left\lbrace \zeta_{i=1,...n} \right\rbrace )=p $  is equal to the number of paths that start from $x_0=0$ and end at $x_{n-1}=p-1$ while satisfying $s_a(\left\lbrace \zeta_{i=1,...n-1} \right\rbrace )=p$, and is given by
\begin{align}
	\begin{aligned}
		&N_n^{(p)}(s_a(\left\lbrace \zeta_i \right\rbrace )=p)
		=
		N_{n-1}^{(p-1)}(s_a(\left\lbrace \zeta_i \right\rbrace )=p)
%		\\
		=
		N_{n-1}^{(p-1)}(s_a(\left\lbrace \zeta_i \right\rbrace ) \geq p)
		-
		N_{n-1}^{(p-1)}(s_a(\left\lbrace \zeta_i \right\rbrace ) \geq p+1)
		\\
		=&
		\dfrac{(n-1)!}{(\frac{n+p}{2})!(\frac{n-p-2}{2})!}
		-
		\dfrac{(n-1)!}{(\frac{n+p+2}{2})!(\frac{n-p-4}{2})!}
%		\\
		=
		(p+2)\dfrac{(n-1)!}{(\frac{n+p+2}{2})!(\frac{n-p-2}{2})!},
	\end{aligned}	
\end{align}
for $n \geq p+4$.
Here in the last equality, we have used Eq.~\eqref{eq:Nnp} while replacing $n$ with $n-1$ and $p$ with $p-1$.
It is straightforward to see $N_n^{(p)}(s_a(\left\lbrace \zeta_i \right\rbrace )=p)$ vanishes for $n=p$, and equals $1$ for $n=p+2$.

We note that, for $n\geq p+4$.
\begin{align}
	\begin{aligned}	
		&\sum_{l=p-1}^{\frac{n+p}{2}}N_n^{(p)}(s_a(\left\lbrace \zeta_i \right\rbrace ) \geq l)
		=\sum_{l=p-1}^{\frac{n+p}{2}} \sum_{k=l}^{\frac{n+p}{2}} N_n^{(p)}(s_a(\left\lbrace \zeta_i \right\rbrace ) = k)
		=
		\sum_{k=p-1}^{\frac{n+p}{2}} (k-p+2) N_n^{(p)}(s_a(\left\lbrace \zeta_i \right\rbrace ) = k)
		\\
		=&
		\sum_{\sum_{i=1}^{n} \zeta_i=p} s_a(\left\lbrace \zeta_i \right\rbrace)
		-(p-2)
		\frac{n!}{(\frac{n+p}{2})!(\frac{n-p}{2})!},
	\end{aligned}	
\end{align}
which leads to
\begin{widetext}
	\begin{align}\label{eq:sa1}
		\begin{aligned}
			&\sum_{\sum_{i=1}^{n} \zeta_i=p} s_a(\left\lbrace \zeta_i \right\rbrace)
			=
			\sum_{l=p+1}^{\frac{n+p}{2}}N_n^{(p)}(s_a(\left\lbrace \zeta_i \right\rbrace ) \geq l)
			+
			N_n^{(p)}(s_a(\left\lbrace \zeta_i \right\rbrace ) \geq p+1)
			+
			N_n^{(p)}(s_a(\left\lbrace \zeta_i \right\rbrace ) = p)
			+
			(p-1)
			\frac{n!}{(\frac{n+p}{2})!(\frac{n-p}{2})!}
			\\
			=&
			\sum_{l=p+1}^{\frac{n+p}{2}}
			\dfrac{n!}{(\frac{n-p}{2}+l)!(\frac{n+p}{2}-l)!}
			+
			\dfrac{n!}{(\frac{n+p}{2}+1)!(\frac{n-p}{2}-1)!}
			+
			(p+2)\dfrac{(n-1)!}{(\frac{n+p+2}{2})!(\frac{n-p-2}{2})!}
			+
			(p-1)
			\frac{n!}{(\frac{n+p}{2})!(\frac{n-p}{2})!}
			\\
%			=&
%			\sum_{l=1}^{\frac{n-p}{2}}
%			\dfrac{n!}{(\frac{n+p}{2}+l)!(\frac{n-p}{2}-l)!}
%			+
%			\dfrac{(n-1)!}{(\frac{n+p}{2}+1)!(\frac{n-p}{2})!}
%			\left[ 
%			n\frac{n-p}{2}
%			+
%			(p+2)\frac{n-p}{2}
%			+
%			(p-1)n\frac{n+p+2}{2}
%			\right] 
%			\\
			=\,&
			\sum_{l=1}^{\frac{n-p}{2}}
			\dfrac{n!}{(\frac{n+p}{2}+l)!(\frac{n-p}{2}-l)!}
			+
			p(n-1)
			\dfrac{(n-1)!}{(\frac{n+p}{2})!(\frac{n-p}{2})!}.	
		\end{aligned}
	\end{align}
\end{widetext}
For $n=p$ and $n=p+2$, $\suml{\sum_{i=1}^{n} \zeta_i=p} s_a(\left\lbrace \zeta_i \right\rbrace)$ takes the value of $p-1$ and $1+p+p^2$, respectively.

Similarly, with the help of the reflection path method, we find the number of paths that begin from $x_0=0$ and end at $x_n=p$ after n steps while satisfying $s_b(\left\lbrace \zeta_i \right\rbrace)=\min \left\lbrace x_i,i=0,...n-1 \right\rbrace \leq -l$ (for $1  \leq l \leq \frac{n-p}{2}$):
\begin{align}
	N_n^{(p)}(\min \left\lbrace x_i \right\rbrace \leq -l)
	=
	\dfrac{n!}{(\frac{n-p}{2}-l)!(\frac{n+p}{2}+l)!}.
\end{align}
This leads to
\begin{align}\label{eq:sb1}
	\sum_{\sum_{i=1}^{n} \zeta_i=p} \!\!\!\!\!\!s_b(\left\lbrace \zeta_i \right\rbrace)
	=
	\begin{dcases}
		-
		\sum_{l=1}^{\frac{n-p}{2}}
		\dfrac{n!}{(\frac{n-p}{2}-l)!(\frac{n+p}{2}+l)!},
		&
		n \geq p+2,
		\\
		0,
		&
		n=p.
	\end{dcases}
\end{align}

Inserting the Eqs.~\eqref{eq:sa1} and~\eqref{eq:sb1} into Eq.~\eqref{eq:Cp}, we find, 
	for odd $p >0$,
	\begin{align}\label{eq:Cp-odd}
		\begin{aligned}
			C_p
			=\,&
			\sum_{m=\frac{p+3}{2}}^{\floor{(N-1)/2}} 
			\frac{1}{2m+1} \frac{1}{2^{2m+2}} 
			\sum_{l=1}^{m+\frac{1-p}{2}} 	2\frac{(2m+1)!}{(m+l+\frac{1+p}{2})!(m-l+\frac{1-p}{2})!}
			+
			p
			\sum_{m=\frac{p+3}{2}}^{\floor{(N-1)/2}} 
			\frac{2m}{2m+1} \frac{1}{2^{2m+2}} 
			\dfrac{(2m)!}{(m+\frac{1+p}{2})!(m+\frac{1-p}{2})!}
			\\
			&+
			\frac{1}{p+2} \frac{1}{2^{p+3}} 
			\left( 
			2+p+p^2
			\right) 
			+ 
			\frac{1}{p} \frac{1}{2^{p+1}} 
			\left( p-1\right) 
			\Theta (p-1),
		\end{aligned}
	\end{align}
	while for even $p>0$
	\begin{align}\label{eq:Cp-even}
		\begin{aligned}
			C_p
			=\,&
			\sum_{m=\frac{p+4}{2}}^{\floor{N/2}} 
			\frac{1}{2m} \frac{1}{2^{2m+1}} 
			\sum_{l=1}^{m-\frac{p}{2}} 2	\frac{(2m)!}{(m+l+\frac{p}{2})!(m-l-\frac{p}{2})!}
			+
			p
			\sum_{m=\frac{p+4}{2}}^{\floor{N/2}} 
			\frac{2m-1}{2m} \frac{1}{2^{2m+1}} 
			\dfrac{(2m-1)!}{(m+\frac{p}{2})!(m-\frac{p}{2})!}
			\\
			&+
			\frac{1}{p+2} \frac{1}{2^{p+3}} 
			\left( 2+p+p^2\right) 
			+
			\frac{1}{p} \frac{1}{2^{p+1}}(p-1).
		\end{aligned}
	\end{align}

We note that the first term in Eq.~\eqref{eq:Cp-odd} is smaller than
\begin{align}
	\sum_{m=0}^{\floor{(N-1)/2}} 
	\frac{1}{2m+1} \frac{1}{2^{2m+2}} 
	2^{2m+1}
	=
	\frac{1}{2} 
	\sum_{m=0}^{\floor{(N-1)/2}} 
	\frac{1}{2m+1} 
	=
	\frac{1}{4}
	\left[ \psi_0(N/2+1)+\gamma_E+\ln 4\right] 
	=
	\frac{1}{4}
	\left( \ln 2 N+\gamma_E \right), 
\end{align}
while the first term in Eq.~\eqref{eq:Cp-even} is smaller than
\begin{align}
	\sum_{m=1}^{\floor{N/2}} 
	\frac{1}{2m} \frac{1}{2^{2m+1}} 
	2^{2m}
	=
	\frac{1}{4}
	\sum_{m=1}^{\floor{N/2} }
	\frac{1}{m} 
	=
	\frac{1}{4}
	\left[ 
	\psi_0(N/2+1)
	+\gamma_E
	\right]  
	=
	\frac{1}{4}
	\left( 
	\ln \frac{N}{2} +\gamma_E
	\right).
\end{align}
The remaining terms in both Eq.~\eqref{eq:Cp-odd} and Eq.~\eqref{eq:Cp-even} converge in the large $N \rightarrow \infty$ limit. As a result, one could draw the conclusion that $C_p$ for $p>0$ is of the order of or smaller compared with $C_0$.

\subsection{F.\ Results}

From previous calculations, we find the SFF $K(t)$ can be expressed as Eq.~\eqref{eq:FT-4}, where $A_p(t)$ admits the form of $A_p(t)
=
-B_p+C_p t$ in the regime of $t \ll N$. 
The coefficient $B_p$ is given by Eq.~\eqref{eq:Bp-1}, while $C_p$ is given by the formulas in Eq.~\eqref{eq:Cp-odd} and Eq.~\eqref{eq:Cp-even} for odd and even positive integer $p$, respectively.
%We emphasize that $C_p$ is of the same order of or smaller than $C_0$ (Eq.~\eqref{eq:C0}).
 
In the regime $t \ll 1$, $ C_p t$ is much smaller compared with $B_p$, and as a result $A_p(t)$ can be approximated by the constant $-B_p$. Combining Eq.~\eqref{eq:Bp-1} and Eq.~\eqref{eq:FT-4}, we find
\begin{align}
\begin{aligned}
	K (t)
	=\,&
	2^N 
	\exp 
	\left\lbrace 
	N \dfrac{J_1(2t)}{t}
	\cos ( \mu t)
	-
	N \ln 2
	-
	2N
	\sum_{p=1}^{N}
	 (-1)^p \frac{1}{p}
	\dfrac{\sin \left(  \pi t p/2  \right)   }{ \pi t p/2  }
	\cos (p \mu t)
	-
	N
	\dfrac{\sin \left(  \pi t /2  \right)   }{ \pi t /2  }
	\cos ( \mu t)
	\right\rbrace 
%	\\
%	=\,&
%	2^N 
%	\exp 
%	\left\lbrace 
%	N \dfrac{J_1(2t)}{t}
%	\cos ( \mu t)
%	-
%	N \ln 2
%	-
%	2N
%	\sum_{p=1}^{N}
%	(-1)^p \frac{1}{p}
%	\cos (p \mu t)
%	-
%	N
%	\cos ( \mu t)
%	\right\rbrace 
	\\
	=\,&
	2^{2N} 
	\exp 
	\left\lbrace 
	N 
	\left( 
	\dfrac{J_1(2t)}{t}
	-
	1
	\right) 
	\cos ( \mu t)
	+
	N
	\ln \left( \frac{1+\cos (\mu t)}{2}\right) 
	\right\rbrace, 
\end{aligned}
\end{align}
where in the second equality, we have approximated $\dfrac{\sin \left(  \pi t p/2  \right)   }{ \pi t p/2  }$ by $1$ for $t \ll 1$.

For $1\ll t \ll N$, $ A_p(t) (-1)^p \dfrac{\sin \left(  \pi t p/2  \right)  }{ \pi t p/2 } \cos(p \mu t)$ is much smaller compared with $A_0(t)$,  and is highly oscillating as a function of $p \geq 1$. Moreover, the overall sign factor $(-1)^p$ is oscillating as well. Therefore, the summation of $ A_p(t) (-1)^p \dfrac{\sin \left(  \pi t p/2  \right)  }{ \pi t p/2 } \cos(p \mu t)$, i.e., the last term in the exponent of Eq.~\eqref{eq:FT-4}, is negligible, and we have
\begin{align}\label{eq:FT-5}
\begin{aligned}
	K(t)
	=
	\exp \left[ 
	N \dfrac{J_1(2t)}{t} \cos (\mu t)
	+
	\frac{t}{4} \left( \ln \frac{N}{8} +\gamma_E \right)
	\right].
\end{aligned}
\end{align}
For time $(N/\ln N)^{2/5} \ll t \ll N$, the first term in the exponent of the above equation is much smaller compared with the second term there [$A_0(t)$], and as a result, $K(t)$ grows as $K(t)=(Ne^{\gamma_E}/8)^{t/4}$ in this regime.
By contrast, for $1 \ll t \ll  (N/\ln N)^{2/5}$, the first term dominates,  and the SFF decays as $K(t)=\exp \left( N \dfrac{J_1(2t)}{t} \cos (\mu t) \right)$.

In summary, using the cluster function approach, we find the many-body SFF $K(t)$ decays rapidly as 
$L^2 \exp	\left\lbrace 
N 
\left( 
\frac{J_1(2t)}{t}
-
1
\right) 
\cos ( \mu t)
+
N
\ln \left(  \frac{1+\cos (\mu t)}{2}\right) 
\right\rbrace $ at early times $t \ll 1$,
and then it continues to drop as
$L\exp\left[  N\dfrac{J_1(2t)}{t}\cos (\mu t)  \right]$ 
for $1  \ll t \ll (N/\ln N)^{2/5}$.
As $t$ becomes larger and lies within the regime $(N/\ln N)^{2/5} \ll t \ll N$, $K(t)$ starts to grow instead of decaying, and exhibits a ramp that scales as $(Ne^{\gamma_E}/8)^{t/4}$. Finally, for $t \gg N$, the SFF should reach the plateau at $K(t)=L$. 
However, it is difficult to extract the behavior for $K(t)$ around $t=N$ due to the presence of unit step function in the expression of $A_p(t)$. In fact, we note that at $t \sim 8 N / \log_2(Ne^{\gamma_E}/8)$, the ramp result $(Ne^{\gamma_E}/8)^{t/4}$ exceeds $L^2$ which is the upper bound for the many-body SFF. Therefore, the expression in Eq.~\eqref{eq:FT-5} is valid only for some $t<\eta_1$ where $\eta_1 < O(N / \log_2 N)$.
In the following section, we use a different approach to study the behavior the many-body SFF $K(t)$ around $t=O(N)$.

\section{Laguerre polynomials  approach}
\label{sec:II}

\subsection{A.\ The SFF in determinant form}

The Wigner-Dyson (WD) distribution for $N$ GUE levels is given by Eq.~\eqref{eq:P_e}. Following Ref.~\cite{refRMTMehta}, we express this in terms of arbitrary polynomials $C_k(x)$ of degree $k-1$ (with arbitrary coefficients, which we will later compensate for by normalizing $P(\eps_1,...,\eps_N)$. Introducing permutation operators $I,J$ that act on $\lbrace 1,...,N\rbrace$ and the symbol $\epsilon_{\{I\}}^{\{J\}} = +1,-1$ if $I,J$ are of the same or opposite parities (respectively), we have
\begin{align}
P(\eps_1,...,\eps_N) &= \left(\prod_i e^{-N\eps_i^2/2}\right) \left(\sum_{I,J}\epsilon_{\{I\}}^{\{J\}}\prod_{k=1}^{N} C_{k}(\eps_{I_k})C_{k}(\eps_{J_k}) \right) \label{eq:polynomialdet} \\
&= \sum_{i,j}\epsilon_{\{i\}}^{\{j\}}\prod_{k=1}^{N} e^{-N\eps_k^2/2}  C_{i_k}(\eps_k)C_{j_k}(\eps_k), \label{eq:newdet}
\end{align}
again up to normalization. The operators $i,j$ in the second line are the inverses of $I,J$ from the corresponding terms in the first line.

It is convenient to choose $C_k(x) = \mathcal{H}_{k-1}\left(\sqrt{\tfrac{N}{2}}x\right)$, where $\mathcal{H}_{k}(x) = (2^k k! \sqrt{\pi})^{-1/2}H_k(x)$ are the normalized Hermite polynomials (satisfying $\int\diff x\ e^{-x^2}\mathcal{H}_i(x) \mathcal{H}_j(x) = \delta_{ij}$). In that case, using Theorem. 5.7.1 in Ref.~\cite{refRMTMehta} to perform the integrals in the normalization condition, 
\begin{equation}
\int \diff \eps_1...\diff \eps_N\ P(\eps_1,...,\eps_N) = 1,
\end{equation}
we can show that the normalized Wigner-Dyson distribution is
\begin{equation}
P(\eps_1,...,\eps_N) = \frac{(N/2)^{N/2}}{N!}\sum_{i,j}\epsilon_{\{i\}}^{\{j\}}\prod_{k=1}^{N} e^{-N\eps_k^2/2}  \mathcal{H}_{i_k-1}\left(\sqrt{\tfrac{N}{2}} \eps_k\right) \mathcal{H}_{j_k-1}\left(\sqrt{\tfrac{N}{2}}\eps_k\right). \label{eq:phermite}
\end{equation}

Inserting Eq.~\eqref{eq:phermite} into Eq.~\eqref{eq:FT-1} and noting that all the integrals over the $\eps_k$ are of the same form, we obtain for the SFF (with the replacement $\eps_k\sqrt{N/2}\to x$ in each factor),
\begin{equation}
K(t) = 2^N\frac{1}{N!}\sum_{i,j}\epsilon_{\{i\}}^{\{j\}}\prod_{k=1}^{N} \int\diff x\ e^{-x^2}  \mathcal{H}_{i_k-1}\left(x\right) \mathcal{H}_{j_k-1}\left(x\right)\left(1+ \cos\left[\left(x\sqrt{\frac{2}{N}}-\mu\right)t\right]\right).
\end{equation}
Using the orthonormality of the $\mathcal{H}_k(x)$, and defining,
\begin{align}
\mdet_{jk}\left(t\right) = \int\diff x\ e^{-x^2} \left\lbrace\mathcal{H}_{j-1}(x)\mathcal{H}_{k-1}(x) \vphantom{\cos\left[\left(x\sqrt{\frac{2}{N}}-\mu\right)t\right]}\cos\left[\left(x\sqrt{\frac{2}{N}}-\mu\right)t\right]\right\rbrace.
 \label{eq:mdef2}
\end{align}
and identifying $(1/N!)\sum_{i,j}\epsilon_{\{i\}}^{\{j\}}A_{ij} = \det A$ for a matrix $A$, we obtain,
\begin{equation}
K(t) = 2^N \det\left[\delta_{jk}+\mdet_{jk}\left(t\right)\right]_{j,k=1,...,N}. \label{eq:SFF_det2}
\end{equation}
Expanding out $\cos((\eps-\mu)T) = \cos(\eps T)\cos(\mu t) + \sin(\eps T)\sin(\mu t)$, the resulting integrals can be evaluated with the help of results from standard tables (e.g. 7.388(6,7) in Ref.~\cite{refGRtables}), and we get
\begin{equation}
\mdet_{j\geq k}(t) = \wdet_{jk}\left(\frac{t^2}{N}\right)\ff_{jk}(\mu t),
\label{eq:mexpr}
\end{equation}
with $\mdet_{kj}(T) = \mdet_{jk}(T)$, where
\begin{equation}
\wdet_{j\geq k}(\tau) = \sqrt{\frac{(k-1)!}{(j-1)!}}\tau^\frac{j-k}{2} e^{-\frac{\tau}{2}} L_{k-1}^{j-k}(\tau),
\label{eq:wdef}
\end{equation}
with $L_n^\alpha(x)$ denoting the Laguerre polynomials, and
\begin{equation}
\ff_{jk}(\mu t) = \begin{cases}
(-1)^{\frac{j-k}{2}}\cos(\mu t), & j-k \text{ is even}, \\
(-1)^{\frac{j-k-1}{2}}\sin(\mu t), & j-k \text{ is odd}.
\end{cases}
\label{eq:fdef}
\end{equation}
Using $\det A = e^{\Tr\ln A}$ in Eq.~\eqref{eq:SFF_det2} and expanding $\ln A$ in a power series, we get
\begin{equation}
K(t) = 2^N \exp\left\lbrace-\sum_{n=1}^{\infty}\frac{(-1)^n}{n}\Tr\ \left[\mdet^n\left(t\right)\right]\right\rbrace.
\label{eq:SFFlogtrace2}
\end{equation}
From Eq.~\eqref{eq:mdef2} and the expression for the cluster functions in Eq.~\eqref{eq:Tn} in terms of Hermite polynomials \cite{refRMTMehta}, it can be shown that this corresponds term-by-term to the cumulant expansion Eq.~\eqref{eq:FT-2}. The expression in terms of Laguerre polynomials however allows us to more easily derive the large-$t$ behavior.

\subsection{B.\ The asymptotic behavior of Laguerre polynomials}
%main text reference
Now we will provide some additional background for the approximate form of Laguerre polynomials $L_n^\alpha(x)$ referred to in the main text. Specifically, we are interested in the form for large $n$, with large or small $\alpha$. Defining $\nu = 4n + 2\alpha+2$, the general behavior of the Laguerre polynomials can be split into the oscillatory region, $x < \nu$, and the monotonic region, $x>\nu$ ~\cite{srefErdelyiLaguerre}. In the oscillatory region, we consider the leading term of Eq.(8) from Sec.10.15 in Ref.~\cite{srefErdelyiLaguerre}, which amounts to
\begin{equation}
e^{-\frac{x}{2}}x^{\frac{\alpha}{2}}L_n^{\alpha}(x)\rvert_{x<\nu} \approx \sqrt{\frac{2}{\pi}}\left(\frac{\nu}{4}\right)^\frac{\alpha}{2}\frac{\sin\left(\varphi_n^\alpha(x)\right)}{(x(\nu-x))^\frac{1}{4}},
\label{eq:Erd_lag_osc}
\end{equation}
where we have absorbed the overall sign into the oscillatory factor $\sin\left(\varphi_n^\alpha(x)\right)$ whose specific form is unimportant for our purposes. In the monotonic region, we have (Eq.(15) form Sec.10.15 in Ref.~\cite{srefErdelyiLaguerre}),
\begin{equation}
e^{-\frac{x}{2}}x^{\frac{\alpha}{2}}L_n^{\alpha}(x)\rvert_{x>\nu} \approx \sqrt{\frac{2}{\pi}}(-1)^n \frac{e^{-\frac{1}{2}\sqrt{x(x-\nu)} + \frac{\nu}{2}\cosh^{-1}\sqrt{\frac{x}{\nu}}}}{(x(x-\nu))^\frac{1}{4}},
\label{eq:Erd_lag_mono}
\end{equation}
which corresponds to a rapid decay to zero. Both of these expressions are valid for $n\to\infty$, but without a corresponding $\alpha\to\infty$ limit.

\begin{figure}
\centering
\includegraphics[scale=0.75]{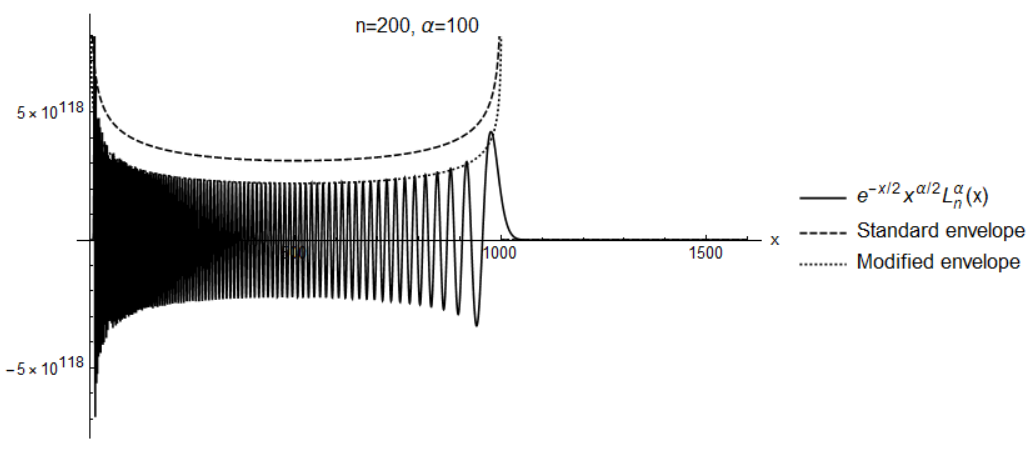} 
\caption{Comparison of approximations for the Laguerre polynomials $L_n^\alpha(x)$ for $n=200, \alpha = 100$. "Standard envelope" refers to Eq.~\eqref{eq:Erd_lag}, and "Modified envelope" to Eq.~\eqref{eq:Laguerre_asymp2}, with the oscillatory factor $\sin\left(\varphi_n^\alpha(x)\right)$ dropped and the overall sign set to $+1$ in both cases.}
\label{fig:aLag}
\end{figure}

For our purposes, we may approximate the right hand side of Eq.~\eqref{eq:Erd_lag_mono} by zero due to the rapid decay of the expression, and cover the entire region of $x<\nu$ and $x>\nu$ by the single approximate expression,
\begin{equation}
e^{-\frac{x}{2}}x^{\frac{\alpha}{2}}L_n^{\alpha}(x) \approx \sqrt{\frac{2}{\pi}}\left(\frac{\nu}{4}\right)^\frac{\alpha}{2}\frac{\sin\left(\varphi_n^\alpha(x)\right)\Theta(\nu-x)}{(x(\nu-x))^\frac{1}{4}}.
\label{eq:Erd_lag}
\end{equation}
We see numerically (cf. Fig.\eqref{fig:aLag}) that Eq.~\eqref{eq:Erd_lag} is not a good approximation for larger $\alpha$ (e.g. $\alpha=O(n)$), given that the $\alpha\to\infty$ limit was not taken in the above standard results. We will need an expression that is valid for any $\alpha\leq n$ for our application. Based on the form of Eq.~\eqref{eq:wdef}, we instead try the approximate expression,
\begin{equation}
e^{-\frac{x}{2}}x^{\frac{\alpha}{2}}L_n^{\alpha}(x) \approx \sqrt{\frac{2}{\pi}\frac{(n+\alpha)!}{n!}}\frac{\sin\left(\varphi_n^\alpha(x)\right) \Theta(\nu-x)}{\left(x(\nu-x)\right)^{\frac{1}{4}}}.
\label{eq:Laguerre_asymp2}
\end{equation}
Numerically, it appears that this expression works better than Eq.~\eqref{eq:Erd_lag} in the desired range of $n,\alpha$ (cf. Fig.\eqref{fig:aLag}), and also has the advantage of greatly simplifying the evaluation of the $n=2$ term in the exponent of Eq.~\eqref{eq:SFFlogtrace2}.

\subsection{C.\ Evaluating the Trace}

Here, we will evaluate the trace in Eq.~\eqref{eq:SFFlogtrace2} for the $n=2$ term, obtaining the late-time ramp. Due to the oscillatory term in Eq.~\eqref{eq:Laguerre_asymp2}, we can approximate all terms with odd $n$ to zero. Also, $\mdet$ is a symmetric matrix, and therefore has real eigenvalues. $\Tr[\mdet^n]$ is then the sum of the $n$-th power of eigenvalues, and for even $n$ must always be positive. While it is hard to precisely evaluate the terms for $n=4,6...$, we can definitively say that their contribution decreases the SFF from the $n=2$ estimate. Therefore, evaluating the $n=2$ term alone would give us an (approximate) upper bound for the SFF.

We have
\begin{equation}
\Tr\left[\mdet^2\left(t\right)\right] = \sum_{j,k=1}^N\mdet^2_{jk}\left(t\right).
\end{equation}
It is convenient to separate out the sum into groups of terms with even or odd $j-k$,
\begin{equation}
\Tr\left[\mdet^2\left(t\right)\right] = \sum_{\substack{j,k=1 \\ \text{even } j-k }}^N\mdet^2_{jk}\left(t\right) + \sum_{\substack{j,k=1 \\ \text{odd } j-k }}^N\mdet^2_{jk}\left(t\right),
\label{eq:mdetsplit}
\end{equation}
As the sums range over $1,...,N$, we expect the indices $j,k$ to typically be large when $N$ is large. Using Eqs.~\eqref{eq:mexpr},~\eqref{eq:wdef},~\eqref{eq:fdef} to write an explicit expression for $\mdet_{jk} = \mdet_{kj}$,
\begin{equation}
\mdet_{j\geq k}(t) = \sqrt{\frac{(k-1)!}{(j-1)!}}\left(\frac{t^2}{N}\right)^\frac{j-k}{2} e^{-\frac{t^2}{2N}} L_{k-1}^{j-k}\left(\frac{t^2}{N}\right) \begin{cases}
(-1)^{\frac{j-k}{2}}\cos(\mu t), & j-k \text{ is even}, \\
(-1)^{\frac{j-k-1}{2}}\sin(\mu t), & j-k \text{ is odd},
\end{cases}
\end{equation}
together with the asymptotic expression for Laguerre polynomials, Eq.~\eqref{eq:Laguerre_asymp2}, gives
\begin{equation}
\mdet_{jk}^2(t) = \frac{2\Theta(\nu_{j+k}-\tau)}{\pi\sqrt{\tau(\nu_{j+k}-\tau)}}\sin^2\left(\varphi_{jk}(\tau)\right)\begin{cases}
\cos^2(\mu t), & j-k \text{ is even}, \\
\sin^2(\mu t), & j-k \text{ is odd},
\end{cases}
\end{equation}
where $\tau = t^2/N$, and $\nu_{j+k}\approx 2(j+k)$ for large $j,k$.

Now, we consider the first sum in Eq.~\eqref{eq:mdetsplit}, with $j-k$ restricted to be even. We transform the summation to the variables $b = (j+k)/2$ (necessarily an integer) and $c=j-k$; $b$ here must be an integer as $j-k$ being even requires that $j+k$ is also even. With $\nu_{j+k} = 4b$, we can write
\begin{align}
\sum_{\substack{j,k=1 \\ \text{even } j-k }}^N\mdet^2_{jk}\left(t\right) &= \sum_{\substack{b,c \\ \text{even } c }}^N \frac{2\Theta(4b-\tau)}{\pi\sqrt{\tau(4b-\tau)}}\sin^2\left(\varphi_{jk}(\tau)\right)\cos^2(\mu t) \\
&\approx \sum_{\substack{b,c \\ \text{even } c }}^N \frac{\Theta(4b-\tau)}{\pi\sqrt{\tau(4b-\tau)}}\cos^2(\mu t),
\end{align}
where in the second line, we have assumed that $\sin^2\left(\varphi_{jk}(\tau)\right)$ oscillates several times over regions of nearly constant $(4b-\tau)^{-1/2}$ (for a given $\tau$) as the latter is typically a slowly varying function (cf. Fig.~\ref{fig:aLag}). This allows us to replace $\sin^2\left(\varphi_{jk}(\tau)\right)$ by its mean value over an oscillation i.e. $1/2$. As the expression is now completely independent of $c$, the sum over $c$ just corresponds to accounting for the number of elements with the same $b$ i.e. the number of elements of an $N\times N$ matrix on the anti-diagonal given by $j+k = 2b$. This number is given by $1+2\min(b-1,N-b)$ as a function of $b$, which we can approximate by $2\min(b, N-b)$ as $N\gg 1$ and $b$ is typically $O(N)$ over most of the sum. We can additionally replace the sum with an integral as the summand/integrand is not a rapidly varying function of $b$, getting
\begin{equation}
\sum_{\substack{j,k=1 \\ \text{even } j-k }}^N\mdet^2_{jk}\left(t\right) \approx \frac{2}{\pi\sqrt{\tau}}\cos^2(\mu t) \left[\int\limits_0^{\frac{N}{2}}\diff b\ \frac{b}{\sqrt{4b-\tau}}\Theta(4b-\tau) + \int\limits_{\frac{N}{2}}^N\diff b\ \frac{N-b}{\sqrt{4b-\tau}}\Theta(4b-\tau)\right].
\end{equation}
These are elementary integrals, and a straightforward calculation gives (as expressed in terms of $t$ rather than $\tau$),
\begin{equation}
\sum_{\substack{j,k=1 \\ \text{even } j-k }}^N\mdet^2_{jk}\left(t\right) \approx \frac{\Theta(2N-t)}{6\pi Nt}\cos^2(\mu t)\left[(4N^2-t^2)^\frac{3}{2}-2(2N^2-t^2)^\frac{3}{2}\Theta(\sqrt{2}N-t)\right].
\end{equation}

For the term with odd $j-k$, almost exactly the same reasoning goes through, except $b$ must now be summed over half-integers (i.e. odd $j+k$), which is a negligible difference when one replaces the sum with an integral in the large $N$ limit. We therefore get
\begin{equation}
\sum_{\substack{j,k=1 \\ \text{odd } j-k }}^N\mdet^2_{jk}\left(t\right) \approx \frac{\Theta(2N-t)}{6\pi Nt}\sin^2(\mu t)\left[(4N^2-t^2)^\frac{3}{2}-2(2N^2-t^2)^\frac{3}{2}\Theta(\sqrt{2}N-t)\right].
\end{equation}
These are to be summed over to obtain the desired trace as in Eq.~\eqref{eq:mdetsplit}. With the trigonometric identity $\cos^2 x+ \sin^2 x =1$ removing all dependence of the expression on the chemical potential $\mu$, we obtain
\begin{equation}
\Tr \left[\mdet^2(t)\right] \approx \frac{\Theta(2N-t)}{6\pi Nt}\left[(4N^2-t^2)^\frac{3}{2}-2(2N^2-t^2)^\frac{3}{2}\Theta(\sqrt{2}N-t)\right],
\end{equation}
leading directly to
\begin{equation}
K(t) \lesssim 2^N \exp\left\lbrace-\frac{\Theta(2N-t)}{12\pi Nt}\left[(4N^2-t^2)^\frac{3}{2}-2(2N^2-t^2)^\frac{3}{2}\Theta(\sqrt{2}N-t)\right]\right\rbrace,
\label{eq:SFFlargeT2}
\end{equation}
We expect this expression to hold for $t>\eta_2=O(N)$ as the approximations we have made i.e. the approximation for Laguerre polynomials in \eqref{eq:Laguerre_asymp2}, and the assumption that our range covers several oscillations of the polynomials, are valid only for $t\sim O(N)$. If we are interested only in $t$ near $t_\ast=2N$, this upper bound also seems to serve as a good approximation to $K(t)$ (observed numerically, as mentioned in the main text), and we get a simple form of the approach of the ramp towards the plateau:
\begin{equation}
K(t) \approx 2^N\exp\left[-\frac{(4N^2-t^2)^\frac{3}{2}}{12\pi Nt}\Theta(2N-t)\right]
\end{equation}
This form is valid in $\sqrt{2}N < t < \infty$, owing to the dropped second term from the exponent of Eq.~\eqref{eq:SFFlargeT2}, and is used as $K_3(t)$ in Eq.(5c) of the main text.

\section{Late-time SFF in the unitary Wigner-Dyson class}
\label{sec:III}
In this section, we comment on the late-time many-body SFF in an arbitrary unitary ensemble of single-particle levels, sharing `local level statistics' with the GUE but with a different single-particle density of states. By not restricting to a Gaussian distribution, we make it highly plausible that any resulting general features would apply universally to the unitary WD class, generalizing the results of Sec.\ref{sec:II}.

The eigenvalue distribution for $N$ levels in an arbitrary unitary ensemble is given by (see e.g. Chap.19 in Ref.\cite{refRMTMehta}),
\begin{equation}
    P(\eps_1,...,\eps_N) \propto \left(\prod_k w(\eps_k)\right) \prod_{1\leq i<j \leq N} \lvert \e_i -\e_j \rvert^2.
\label{eq:WDgeneral}
\end{equation}
The quadratic factors $\lvert\e_i-\e_j\rvert^2$ contain the universal WD level repulsion for the unitary class, while the weights $w(\eps)$ determine the overall distribution of states. We will assume that $w(\eps)$ is strictly positive in a connected interval, and zero everywhere else if this interval is finite. The special case of Gaussian weights gives the GUE distribution, Eq.~\eqref{eq:P_e}. We can write the eigenvalue distribution in terms of arbitrary polynomials $C_k(x)$ of degree $k-1$ in a manner completely analogous to Eq.~\eqref{eq:polynomialdet}, obtaining
\begin{equation}
    P(\eps_1,...,\eps_N) \propto \left(\prod_k w(\eps_k)\right) \left(\sum_{I,J}\epsilon_{\{I\}}^{\{J\}}\prod_{k=1}^{N} C_{k}(\eps_{I_k})C_{k}(\eps_{J_k}) \right).
\end{equation}
Instead of choosing Hermite polynomials, it is now convenient to choose polynomials $f_n(x)$ that are orthonormal with respect to the weight $w(x)$, i.e.
\begin{equation}
    \int\diff x\ w(x) f_n(x)f_n(x) = \delta_{jk}.
\end{equation}
This can always be done, for instance, by starting with the functions $\lbrace x^k\rbrace_{k=0}^{\infty}$ and applying the Gram-Schmidt orthonormalization procedure, with the inner product of functions defined with the weight $w(x)$ in the connected interval of nonzero $w(x)$.

To calculate the SFF of the corresponding many-body system, we set the explicit chemical potential $\mu$ to zero as it can now be absorbed into the mean of $w(x)$, which we have allowed to be arbitrary. We can then largely imitate the derivation of Eq.~\eqref{eq:SFFlogtrace2} for the many-body SFF, replacing the Hermite polynomials with $f_n(x)$ and obtaining
\begin{equation}
K(t) = 2^N \exp\left\lbrace-\sum_{n=1}^{\infty}\frac{(-1)^n}{n}\Tr\ \left[\Lambda^n\left(t\right)\right]\right\rbrace,
\label{eq:SFFlogtracegeneral}
\end{equation}
where we have defined the matrix $\Lambda$ in analogy with $\mdet$ as
\begin{equation}
    \Lambda_{jk}(t) = \int\diff x\ w(x)f_j(x)f_k(x) \cos(xt).
    \label{eq:Lambdadef}
\end{equation}
In what follows, 'large $t$' or 'late times' in this section will always refer to when $t$ is at the order of magnitude of (the inverse of) the scale of single-particle level spacings, as opposed to (the inverse of) the scale of variation of the single-particle density of states. The latter of these is typically smaller by a factor of $\sim N$ in the time domain.

As with Eq.~\eqref{eq:SFFlogtrace2}, Eq.~\eqref{eq:SFFlogtracegeneral} can again be identified term-by-term with the cumulant expansion (analogous to Eq.~\eqref{eq:FT-2}) for the ensemble, which will be useful to keep in mind for subsequent calculations. We assume that at late times ($t$ of the order of single-particle level spacings), there is negligible contribution from terms with odd values of $n$, leaving us with a sum over only even values of $n$ in Eq.~\eqref{eq:SFFlogtracegeneral}. The vanishing of the $n=1$ term at late times is immediate as it corresponds to the Fourier transform of the slowly varying single-particle density of states (e.g.~\cite{refRMTMehta}), and we will justify this assumption for odd $n\geq 3$ later in this section. On account of $\Lambda_{jk}$ being a symmetric matrix, the traces with even $n$ are guaranteed to be positive, so retaining any one term in the sum and dropping all the other terms gives us an upper bound for the SFF.

We again choose to focus on the $n=2$ term for its simplicity, in which the trace can be written as
\begin{equation}
    \Tr[\Lambda^2(t)] = \int\diff x\int \diff y\ w(x)w(y) \left(\sum_j f_j(x)f_j(y)\right) \left(\sum_k f_k(x)f_k(y)\right) \cos(xt) \cos(yt).
\end{equation}
We recognize the RMT kernel~\cite{refRMTMehta}
\begin{equation}
    \kk(x,y) = \sum \sqrt{w(x)w(y)}\sum_k f_k(x)f_k(y),
    \label{eq:RMTkerneldef}
\end{equation}
and note that it is an even function. Determinants constructed from $\kk(x,y)$ with the appropriate arguments give all the single-particle correlation functions $\R_n$ of the ensemble, as in Eq.~\eqref{eq:Rn}. We then obtain
\begin{align}
    \Tr[\Lambda^2(t)] &= \int\diff x\int \diff y\ \kk(x,y)\kk(y,x) \cos(xt) \cos(yt) \nonumber \\
    &= \frac{1}{2}\int\diff x\int \diff y\ \kk(x,y)\kk(y,x) \cos((x+y)t) +\frac{1}{2}\int\diff x\int \diff y\ \kk(x,y)\kk(y,x) \cos((x-y)t). \label{eq:L2sum}
\end{align}

For any unitary ensemble with the above assumptions for $w(\eps)$, it appears that $\kk(x,y)$ has the universal form (for $N\to\infty$) \cite{refRMTMehta},
\begin{equation}
    \kk(x,y) \approx\frac{\sin(\rho(x)\pi(x-y))}{\pi(x-y)},
    \label{eq:kkgendef}
\end{equation}
where $\rho(x)$ is the single-particle density of states, which is ensemble-specific. Eq.~\eqref{eq:kkgendef} should ideally symmetrize $\rho(x)$ in $x$ and $y$, but typically $\rho(x)\approx\rho(y)$ if $x$ and $y$ are close enough for $\kk(x,y)$ to be non-negligible (i.e. are comparable to the scale of single-particle level spacings) and we can ignore the symmetrization. The slow variation of $\rho(x)$ at this scale also implies that for large $t$, $\kk(x,y)$ has a negligible Fourier component with respect to $x+y$ or either of $x$ or $y$, for fixed $x-y$. This allows us to neglect the first term in Eq.~\eqref{eq:L2sum}. We can idenfify the second term in Eq.~\eqref{eq:L2sum} as essentially the connected part of the single particle SFF,
\begin{equation}
    \kappa_c(t) = -\int\diff x\int \diff y\ \kk(x,y)\kk(y,x) \cos((x-y)t),
    \label{eq:kcdef}
\end{equation}
which is related to the full single particle SFF $\kappa(t)$ by
\begin{align}
    \kappa(t) &= \int\diff \eps_1\int\diff\eps_2\ \R_2(\eps_1,\eps_2)\cos((\eps_1-\eps_2)t) \nonumber \\
    &= N + \lvert\tilde{\rho}(t)\rvert^2 + \kappa_c(t),
\end{align}
where $\tilde{\rho}(t)$ is the Fourier transform of the single-particle density of states. Using this identification in the $n=2$ term of Eq.\eqref{eq:SFFlogtracegeneral} leads to the inequality
\begin{equation}
    K(t) \leq 2^N \exp\left[\frac{\kappa_c(t)}{4}\right] \to 2^N\exp\left[\frac{\kappa(t)-N}{4}\right],
    \label{eq:sffbound2}
\end{equation}
where we have assumed that $\tilde{\rho}(t)$ vanishes at the large times of interest, which follows from the slow variation of $\rho(x)$. This inequality generalizes Eq.~\eqref{eq:SFFlargeT2} to non-Gaussian ensembles.

We will also show that the many-body SFF has the same plateau time $t'_\ast$ as the single-particle SFF, and simultaneously establish that all terms with odd $n$ in Eq.~\eqref{eq:SFFlogtracegeneral} vanish at late times. The approach used here has similarities to the computation of the GUE cumulant expansion in Sec.~\ref{sec:tm}. We consider the nonvanishing higher order terms in Eq.~\eqref{eq:SFFlogtracegeneral}, expressed in terms of the RMT kernel in Eq.~\eqref{eq:RMTkerneldef} (with $n\geq 3$),
\begin{equation}
    \Tr[\Lambda^{n}(t)] = \int\diff^{n}x\  \kk(x_1,x_2)\kk(x_2,x_3)...\kk(x_{n}, x_1) \prod_{j = 1}^{n} \cos(x_j t),
    \label{eq:higherlambda}
\end{equation}
with $\diff^{n}x = \prod_{i=1}^{n}\diff x_i$. Let $\tilde{\kk}(\tau_i, x_i)$ be the Fourier transform of $\kk(x_i,x_{i+1})$ (we adopt the convention of a periodic subscript i.e. $x_{n+1}=x_1$ and $x_{0} = x_{n}$) with respect to the difference $(x_{i}-x_{i+1})$, which still depends on $x_i$ (as argued after Eq.~\eqref{eq:kkgendef}, $x_i$ is interchangeable with $x_{i+1}$ here). We can then write,
\begin{equation}
\kk(x_i,x_{i+1}) = \int\frac{\diff\tau_i}{2\pi}\ \tilde{\kk}(\tau_i, x_i)\cos\left(\left(x_i-x_{i+1}\right)\tau_i\right).
\label{eq:kkgenfourier}
\end{equation}
It is important to emphasize that, as with $\kk(x_i, x_{i+1})$, $\tilde{\kk}(\tau_i, x_i)$ does not have a high frequency/large-$t$ Fourier component in $x_i$ - all the dependence on $x_i-x_{i+1}$ is contained in $\tau_i$.

Inserting Eq.~\eqref{eq:kkgenfourier} in Eq.~\eqref{eq:higherlambda} gives,
\begin{equation}
\Tr[\Lambda^{n}(t)] = \int\diff^{n}x\int \frac{\diff^{n}\tau}{(2\pi)^{n}}\ \left(\prod_{i=1}^{n}\tilde{\kk}(\tau_i, x_i)\right) \left[\left\lbrace \prod_{l=1}^{2k}\cos\left(x_l\tau_l-x_{l+1}\tau_l\right)\right\rbrace \prod_{j = 1}^{n} \cos(x_j t)\right].   
\label{eq:LambdaFourier1}
\end{equation}
The product of all the different cosine factors can be expanded into a sum over cosine terms using standard trigonometry,
\begin{equation}
    \prod_{i,j=1}^{n} \cos\left(\left(x_i-x_{i+1}\right)\tau_i\right)\cos(x_j t) = \sum_{\lbrace\zeta_i,s_i = \pm 1\rbrace} 2^{-2n}\cos\left[\sum_{j=1}^n\left(\zeta_j t +s_j \tau_j - s_{j-1}\tau_{j-1}\right)x_j\right].
\label{eq:cos_ptos}
\end{equation}
We can now further expand each of these cosine terms, again using standard trigonometric identities, as a sum of products of sines and cosines of each $\left[\left(\zeta_j t +s_j \tau_j - s_{j-1}\tau_{j-1}\right)x_j\right]$,
\begin{equation}
\cos\left[\sum_{j=1}^n\left(\zeta_j t +s_j \tau_j - s_{j-1}\tau_{j-1}\right)x_j\right] = \sum_{\lbrace \ell_k = \pm 1\rbrace} \mathcal{C}_{\ell_1...\ell_n} \prod_{j=1}^{n}\cs_{\ell_j}\left[\left(\zeta_j t +s_j \tau_j - s_{j-1}\tau_{j-1}\right)x_j\right],
\label{eq:cos_expansion}
\end{equation}
where we have introduced notation for a general sinusoid, $\cs_{+1}(x) = \cos(x)$ and $\cs_{-1}(x) = \sin(x)$, and the coefficients satisfy $\mathcal{C}_{\ell_1...\ell_n} \in \lbrace-1,0,1\rbrace$ (their exact values are unimportant for our purposes). Combining Eqs.~\eqref{eq:LambdaFourier1},~\eqref{eq:cos_ptos} and \eqref{eq:cos_expansion}, and changing the order of sums and integration over $x$ and $\tau$ gives
\begin{equation}
\Tr[\Lambda^{n}(t)] = \sum_{\lbrace\zeta_i,s_i, \ell_i = \pm 1\rbrace} 2^{-2n}\mathcal{C}_{\ell_1...\ell_n} \int \frac{\diff^{n}\tau}{(2\pi)^{n}} \left(\prod_{j=1}^n \int\diff x_j\ \tilde{\kk}(\tau_i, x_i) \cs_{\ell_j}\left[\left(\zeta_j t +s_j \tau_j - s_{j-1}\tau_{j-1}\right)x_j\right] \right).
\label{eq:Lambdasinusoid}
\end{equation}
 
Each term in the $\tau$-integrand of Eq.~\eqref{eq:Lambdasinusoid} has factorized neatly into a product of Fourier cosine or sine transforms of $\tilde{\kk}(\tau_i, x_i)$ with respect to $x_i$. We have already noted that this Fourier component exists only at low frequencies - unless the coefficient of $x_j$ in the argument of the sinusoid ($\cs_{\ell_j}$) is comparable to the small time scales (say, $\sim t_s$) where $\tilde{\rho}(t)$ is appreciable, the integral vanishes due to the rapidly oscillating cosine. But as $t_s$ is typically much smaller than the plateau time $t'_\ast$ by a factor of $\sim N$, we can approximate $t_s \sim 0$ at large times, requiring the argument of each sinusoid to vanish. We then obtain the condition,
\begin{equation}
    \zeta_j t +s_j \tau_j - s_{j-1}\tau_{j-1} \approx 0, \ \forall j,
    \label{eq:tcondition}
\end{equation}
for at least one choice of the sign coefficients $\lbrace \zeta_i,s_i\rbrace$, as a necessary condition for a nonzero value for $\Tr[\Lambda^{n}(t)]$. Adding up Eq.\eqref{eq:tcondition} over all values of $j$ gives the constraint $\sum_{j=1}^n \zeta_j = 0$ in non-vanishing terms. This condition (valid for $n\geq 3$) is impossible to satisfy for odd $n$, and together with the vanishing of the $n=1$ contribution, justifies the assumption made earlier that we can ignore the contribution from all odd values of $n$ at late times.

To obtain the plateau time, we note that substituting Eq.~\eqref{eq:kkgenfourier} in Eq.~\eqref{eq:kcdef} gives by the standard Fourier convolution relation (treating $x$ and $y$ as interchangeable),
\begin{equation}
 \kappa_c(t) = \int\diff x\int\frac{\diff\tau}{2\pi}\tilde{\kk}(t-\tau, x)\tilde{\kk}(\tau, x). 
\end{equation}
We know that $\kappa_c(\lvert t\rvert>t'_\ast) = 0$, corresponding to the plateau (with both positive and negative values of $t$), which requires that $\kk(\tau > t'_\ast/2, s) = 0$. This means that Eq.~\eqref{eq:tcondition} can be satisfied only if $\lvert t\rvert < t'_\ast$, which implies
\begin{equation}
    \Tr[\Lambda^n(t>t'_\ast)] = 0.
\label{eq:lambdaplateau}
\end{equation}

In summary, Eq.~\eqref{eq:sffbound2} and Eq.~\eqref{eq:lambdaplateau} with Eq.~\eqref{eq:SFFlogtracegeneral}, taken together, provide the following universal conditions that apply to the late-time many-body SFF of any unitary Wigner-Dyson ensemble:
\begin{align}
    K(t\leq t'_\ast) &\leq 2^N \exp\left[\frac{\kappa(t)-N}{4}\right], \\
    K(t>t'_\ast) &= 2^N,
\end{align}
where $\kappa(t)$ is the single-particle SFF with plateau time $t'_\ast$.

\section{$\sigma$-model approach}
\label{sec:IV}

In this section, we introduce a $\sigma$-model approach to investigate the SFF $K(t)$ for the ensemble of noninteracting GUE Hamiltonian (Eqs.~\eqref{eq:H} and~\eqref{eq:P_h}). We construct a path integral formula for $K(t)$ and connect it with the cumulant expansion discussed in Sec.~\ref{sec:I}. Then we derive a $\sigma$-model, from which we argue that the SFF can be recovered by performing a non-perturbative summation. This $\sigma$-model approach can be directly generalized to the interacting case, and is therefore especially useful for the investigation of the structure of the many-body energy levels for many-body chaotic systems.

The starting point is the path integral formula for the SFF:
\begin{align}\label{eq:ZT}
\begin{aligned}
K(t)
=\,&
\left\langle 
\int \D (\bpsi, \psi) 
\exp 
\left\lbrace 
\suml{a,b=L,R}
i
\int_0^{t} d t'
\bpsi_i^{a} (t')
\left[ 
(i \partial_{t'} +\mu \sigma^3 )\delta_{ij}
-
h_{ij} \sigma^3
\right]_{ab}
\bpsi_j^{b} (t')
\right\rbrace
\right\rangle.	
\end{aligned}	
\end{align}
Here the integrals over $\psi^{L}$ and $\psi^{R}$, respectively, lead to $Z(it)$ and $Z(-it)$.  $\sigma$ indicates the Pauli matrix in the $L/R$ space.
The fermionic field $\psi$ is subject to the boundary condition
$
\psi(t)=-\psi(0).
$
After the Fourier transform,
the above equation reduces to
\begin{align}\label{eq:Zn}
\begin{aligned}
K(t)
=
\left\langle 
\int 
\D (\bpsi, \psi)
\exp 
\left\lbrace 
i \suml{n}
\bpsi_{i,n}^{a} 
\left[ 
(\ww_n +\mu \sigma^3 )\delta_{ij}
-
h_{ij} \sigma^3
\right]_{ab}
\psi_{j,n}^{b} 	
\right\rbrace
\right\rangle,
\end{aligned}	
\end{align}
where the subscript $n$ represents the Matsubara frequency $\ww_n=\frac{2\pi}{t} (n+\frac{1}{2})$, and we have employed the convention that repeated indices imply summation.

\subsection{A.\ Connection with Cumulant Expansion }

Before moving to the derivation of $\sigma$-model, in this section we integrate out the fermionic field first before ensemble averaging, in an attempt to see the connection between the current approach and cluster function approach described in Sec.~\ref{sec:I}.
After integrating out the fermionic field $\psi$ in Eq.~\eqref{eq:Zn}, we have
\begin{align}\label{eq:Z1}
\begin{aligned}
K(t)
%	=&
%	\left\langle 
%	\exp \left\lbrace  
%	\sum_n \Tr \ln \left\lbrace - i t   \left[ ( \ww_n + \mu )\delta_{ij}- h_{ij}\right] \right\rbrace 
%	+
%	\sum_n \Tr \ln \left\lbrace  i t   \left[ (- \ww_n+ \mu )\delta_{ij}- h_{ij}\right]
%	\right\rbrace  
%	\right\rbrace 
%	\right\rangle 
%	\\
=&
\left\langle 
\exp 
\left\lbrace  
\suml{n} \suml{i=1}^{N} 
\left\lbrace
\ln \left[- i t  \left( 	\ww_n + \mu - \e_i \right) \right] 
+
\ln \left[  i t  \left( -	\ww_n  + \mu - \e_i \right)\right]  
\right\rbrace 
\right\rbrace 
\right\rangle,
\end{aligned}
\end{align}
where $ \e_i $ is the eigenvalue of the Hermitian matrix $h$.
Carrying out the summation over the Matsubara frequency $\ww_n=\frac{2\pi}{t} (n+\frac{1}{2})$, we find the above equation reduces to Eq.~\eqref{eq:FT-0}, which is the starting point of the cluster function approach.  It can be rewritten as
\begin{align}\label{eq:Kave}
\begin{aligned}
K(t)
=
\left\langle
\exp \left[ 	  
N
\int_{-\infty}^{\infty} d \ww
\ln \left[ 2+ 2\cos \left( t(\ww-\mu) \right) \right] 
\nu(\ww)
\right] 
\right\rangle,
\end{aligned}
\end{align}
where $
\nu(\ww)=\frac{1}{N} \sum_{i=1}^N \delta (\ww - \e_i)
$
is the density of states (DOS).
We then introduce a new correlation function of the single-particle energy levels:
\begin{align}\label{eq:TR}
\begin{aligned}
\R_n'(\ww_1,...,\ww_n)
=
N^n	
\left\langle 
\nu(\ww_1)...\nu(\ww_n)
\right\rangle,
\end{aligned}
\end{align}
and the corresponding cluster function $\T_n'(\ww_1,...,\ww_n)$.
The relationship between $\R_n'$ and $\T_n'$ is equivalent to that between $\R_n$ and $\T_n$ in Eq.~\eqref{eq:cluster}.
We emphasize that $\R_n'$ is defined slightly differently from $\R_n$ [Eq.~(\ref{eq:Rn})] and contains several extra $\delta$-functions. 

A cumulant expansion of Eq.~\eqref{eq:Kave} leads to
\begin{align}\label{eq:CumSum}
\begin{aligned}
&\ln K(t)  
=\,
\int_{-\infty}^{\infty} d \ww
\ln \left[ 2+ 2\cos \left( t(\ww-\mu) \right) \right] 
\T_1'(\ww)
\\
-&
\frac{1}{2} 
\int_{-\infty}^{\infty} d \ww
\int_{-\infty}^{\infty} d \ww'
\ln \left[ 2+ 2\cos \left( t(\ww-\mu) \right) \right] 
\ln \left[ 2+ 2\cos \left( t(\ww'-\mu) \right) \right] 
\T_2'(\ww,\ww')+...
\\
+&
(-1)^{n-1}
\frac{1}{n!} 
\int_{-\infty}^{\infty} ...\int_{-\infty}^{\infty} 
\prod_{k=1}^{n}
\left\lbrace 
d \ww_k
\ln \left[ 2+ 2\cos \left( t(\ww_k-\mu) \right) \right] 
\right\rbrace 
\T_n'(\ww_1,...,\ww_n)
+... \,\,\,.
\end{aligned}
\end{align}
By connecting $\T_n'$ and $\T_n$, this equation can be proven to be equivalent to the previous cumulant expansion Eq.~\eqref{eq:FT-2} utilized in Sec.~\ref{sec:I} .

\subsection{B. $\sigma$-model for the SFF}

The ensemble average in Eq.~\eqref{eq:Zn} can be re-expressed as
\begin{align}\label{eq:aveK}
\begin{aligned}
&K(t)
=
\frac{1}{\int \D h e^{-\frac{N}{2} \Tr h^2 } }
{
	\int \D h
	e^{-\frac{N}{2} \Tr h^2 }
	\int 
	\D (\bpsi, \psi)
	\exp 
	\left\lbrace 
	i \suml{n}
	\bpsi_{n}
	\left[ 
	(\ww_n +\mu \sigma^3 )
	-
	h \sigma^3
	\right]
	\psi_{n}
	\right\rbrace
}
\\
=&
\int \D (\bpsi, \psi)
\exp 
\left\lbrace 
\begin{aligned}
i \suml{n}
\bpsi_{n}
\left[ 
\ww_n+\mu\sigma^3
\right]	
\psi_{n}
- 
\frac{1}{2N}  \bpsi_{j,m}\sigma^3 \psi_{i,m} \bpsi_{i,n} \sigma^3 \psi_{j,n}
\end{aligned}	
\right\rbrace
\\
=&
\frac{1}{\int \D Q e^{-\frac{N}{2}\Tr Q^2} }
\int \D Q e^{-\frac{N}{2}\Tr Q^2}
\int \D (\bpsi, \psi)
\exp 
\left\lbrace 
\begin{aligned}
i \suml{n,m}
\bpsi_{i,n}
\left\lbrace 
\left[ 
(\ww_n\sigma^3 +\mu )\delta_{nm}
-
i Q_{nm}
\right]
\sigma^3
\right\rbrace 
\psi_{i,m}
\end{aligned}	
\right\rbrace.
\end{aligned}	
\end{align}
In the 2nd equality, we have integrated over $h$ first, resulting in a quartic interaction term.
Then, in the 3rd equality, this quartic term is decoupled by a Hermitian matrix field $Q$.  Integrating over the fermionic field $\psi$ in Eq.~\eqref{eq:aveK}, we obtain the $\sigma$-model:
\begin{align}\label{eq:sigmaM}
\begin{aligned}
&K(t)=\,\dfrac{\int \D Q \exp \left( - S[Q]\right) }
{\int \D Q e^{-\frac{N}{2}\Tr Q^2}  },
\qquad
S[Q]=\,	
\frac{N}{2}
\Tr Q^2
-
N\Tr \ln
\left\lbrace 
-t
\left[
i(\hat{\ww} \sigma^3 +\mu  )
+
Q
\right]	
\sigma^3
\right\rbrace,
\end{aligned}
\end{align}
where $\hat{\ww}$ represents a diagonal matrix with elements given by $\hat{\ww}^{ab}_{nm}=\delta_{ab}\delta_{nm}\ww_n$.

We note that this $\sigma$-model is largely similar to the one derived by Kamenev and M{\'e}zard in Ref.~\cite{Kamenev-GUE}, from which they recovered the semi-circle law for the averaged single-particle level density $\R_1(\ww)$ as well as the two-point correlation function of single-particle levels $\R_2(\ww,\ww')$ by considering the contribution from the saddle point and the quadratic order fluctuations around it. 
%A generalized version of the current $\sigma$-model can serve as a starting point for investigating the many-body level statics of interacting model. The detailed calculation will be present elsewhere.

\subsection{C. Saddle Point and Fluctuations}

In the following, we will describe the zero, soft and massive modes from the fluctuations around the saddle point and discuss briefly their contribution to the SFF.
The saddle point equation can be obtained by taking the variation of the action $S[Q]$ with respect to $Q$:
\begin{align}
\begin{aligned}
Q_{sp}
=
\left[
i(\hat{\ww} \sigma^3 +\mu  )
+
Q_{sp}
\right]^{-1},
\end{aligned}
\end{align}
and its solution can be expressed as $Q_{sp}=U^{-1} \Lambda U$. Here $\Lambda$ is a diagonal matrix with element 
\begin{align}\label{eq:Lambda}
\begin{aligned}
\Lambda^{ab}_{nm}
=
\begin{cases}
\frac{i}{2}
\left[ -(\zeta_a\ww_n + \mu)+ \sgn (\zeta_a\ww_n + \mu)\sqrt{(\zeta_a \ww_n + \mu)^2-4} \right] 
\delta_{ab}\delta_{nm},
&
|\zeta_a \ww_n + \mu|>2,
\\
\frac{1}{2}
\left[ -i(\zeta_a\ww_n + \mu)+ s_n^a \sqrt{4-(\zeta_a \ww_n + \mu)^2} \right] 
\delta_{ab}\delta_{nm},
&
|\zeta_a \ww_n + \mu| \leq 2,
\end{cases}
\end{aligned}
\end{align}
where $\zeta_{L/R}=\pm 1$. $s_n^a$ can take the value of $+1$ or $-1$, resulting in various diagonal matrices $\Lambda^{(s)}$ (see Ref.~\cite{Kamenev-GUE}) which are essential for the calculation of the SFF.
$U$ is a directly product of multiple $\frac{U(2)}{U(1) \times U(1)}$ rotation matrices, each of which applies in the space of $L,\ww_n$ and $R,-\ww_n$. In other words, $U^{ab}_{nm}$ is nonzero only when $m=-n$ and $a\neq b$, or $n=m$ and $a=b$.

We then consider fluctuations around the saddle point. We express $Q$ as $Q=Q_{sp}+U^{-1} \delta Q U$ and insert it into the action $S[Q]$ (Eq.~\eqref{eq:sigmaM}). Expanding in terms of $\delta Q$ up to the quadratic order leads to
\begin{align}\label{eq:S2}
\begin{aligned}
&K(t)
=\,
\suml{s}
e^{-S[\Lambda^{(s)}]}
\dfrac{\int \D \delta Q \exp \left( 
	-\frac{N}{2}\suml{abnm}
	\delta Q_{nm}^{ab}
	\left[  
	1+
	G_a(\ww_n) G_b(\ww_m)
	\right] 
	\delta Q_{mn}^{ba}
	\right) }
{\int \D Q e^ { -\frac{N}{2} 	\Tr Q^2} },
%	\\
%	&S[\Lambda^{(s)}]
%	=
%	N
%	\sum_{n}
%	\left\lbrace 
%	\frac{(\Lambda^{LL}_{nn})^2 +(\Lambda^{RR}_{nn})^2}{2}
%	-
%	\ln \left\lbrace -t\left[ i(\ww_n +\mu)+\Lambda^{LL}_{nn}\right] \right\rbrace 
%	-
%	\ln \left\lbrace t\left[i(-\ww_n + \mu)+\Lambda^{RR}_{nn}\right] \right\rbrace 
%	\right\rbrace, 
%\\
%&
%S^{(2)}[\delta Q]
%=
%\frac{1}{2}\suml{abnm}
%\delta Q_{nm}^{ab}
%\left[  
%1+
%G_a(\ww_n) G_b(\ww_m)
%\right] 
%\delta Q_{mn}^{ba}.
\end{aligned}
\end{align}
where
\begin{align}
\begin{aligned}
&G_a(\ww_n)
\equiv	
\left[ 
i(\zeta_a \ww_n +\mu)
+
(\Lambda^{(s)})^{aa}_{nn}
\right]^{-1},
%	=
%	2 \left[ i (\zeta_a \ww_n + \mu)+s^{a}_n \sqrt{4-(\zeta_a \ww_n + \mu)^2} \right]^{-1},
\end{aligned}
\end{align}
and $\sum_{s}$ represents a summation over various diagonal saddle points $\Lambda^{(s)}$ (Eq.~\eqref{eq:Lambda}) which are not connected by the rotation $\Lambda^{(s')}=U^{-1}\Lambda^{(s)}U$.

We find various modes for $\delta Q_{nm}^{ab}$ depending on whether $s_n^a$ is equal or opposite to $s_m^b$. 
We focus on the case where both $|\zeta_a \ww_n + \mu|$ and $|\zeta_b \ww_m + \mu|$ are smaller than $2$.
When $s_n^a=-s_m^b$, the kernel of the quadratic action $1+ G_a(\ww_n) G_b(\ww_m)$ is massless, meaning it vanishes when $\zeta_a \ww_n = \zeta_b\ww_m$.
For small but nonzero $\zeta_a \ww_n - \zeta_b\ww_m$, the kernel can be approximated by
\begin{align}
\begin{aligned}
1+G_a(\ww_n) G_b(\ww_m)
=
\frac{-is_n^a(\zeta_b\ww_m-\zeta_a\ww_n)}{\sqrt{4-(\zeta_a\ww_n+\mu)^2}}
+
O((\zeta_b\ww_m-\zeta_a\ww_n)^2).
\end{aligned}
\end{align}
These correspond to the soft modes.
For the extreme case when $\zeta_a \ww_n =\zeta_b\ww_m$, $1+G_a(\ww_n) G_b(\ww_m)$ vanishes and the associated modes are the zero modes.
By contrast, when $s_n^a=s_m^b$, the kernel has a mass. It remains nonzero when $\zeta_a \ww_n=\zeta_b\ww_m$, and is approximately
\begin{align}
\begin{aligned}
1+G_a(\ww_n) G_b(\ww_m)
=
1+\frac{4}{\left[ i(\zeta_a\ww_n+\mu)+s_n^a\sqrt{4-(\zeta_a\ww_n+\mu)^2} \right]^2}
+
O(\zeta_b\ww_m-\zeta_a\ww_n).
\end{aligned}
\end{align}
for small $\zeta_b\ww_m-\zeta_a\ww_n$.
These modes are the massive modes.

Integrating over $\delta Q$ in Eq.~\ref{eq:S2}, one can see that, at the quadratic level, each mode contributes to the SFF a factor of $1/\sqrt{1+G_a(\ww_n) G_b(\ww_m)}$ which becomes divergent for the zero modes. Higher order fluctuations are therefore required to regularize this infrared divergence. Integration over the zero mode yields a factor proportional to the volume of the saddle point manifold (see discussion around Eq.~31 in Ref.~\cite{Kamenev} and the Appendix therein). With proper rescaling which accounts for the normalization constant, i.e., the denominator in Eq.~\eqref{eq:sigmaM}, one can see that the zero modes give rise to a factor of $N^{\text{const.} \times t}$, reproducing a behaviour similar to Eq.~\eqref{eq:FT-5} obtained by the cluster function approach (Sec.~\ref{sec:I}).
The contribution from the remaining soft modes and massive modes is comparable, and is important to determine the coefficient in the exponent of the $N^{\text{const.} \times t}$ ramp.
To see that, one may need to consider fluctuations beyond the quadratic order, and a calculation similar to that in Refs.~\cite{Kamenev-GUE} and~\cite{Kamenev} can reproduce each term one by one in the cumulant expansion Eq.~\eqref{eq:CumSum}  for the nonzero soft modes and massive modes. 
We note that a summation over $N\rightarrow \infty$ terms as in the cluster function approach (Sec.~\ref{sec:I}) is also required to find the total contribution from nonzero modes. 

One of the advantages of this $\sigma$-model approach is that the level correlation functions can be obtained automatically, while in the previous approach we directly employ Mehta's result~\cite{refRMTMehta}. Most importantly, this $\sigma$-model approach can be generalized to the interacting case, and is more suitable for investigating the many-body level statics of interacting models. In the presence of interactions, we expect that the soft modes acquire a mass term, which results in a drastically different behavior of the SFF. The detailed calculation will be presented in a separate study.

\bibliography{main}